\documentclass[aps,pra,twocolumn,showpacs,letterpaper,tighten,float,superscriptaddress,footinbib]{revtex4-2}
\usepackage{amssymb}
\usepackage{amsbsy}
\usepackage{amsmath}
\usepackage{mathrsfs}
\usepackage{epsfig}
\usepackage{graphicx}
\usepackage{array}
\usepackage{textcomp}
\usepackage{color}
\usepackage{braket}
\usepackage{bm,hyperref}
\usepackage[justification=raggedright,font=small]{caption}
\usepackage[titletoc,toc,title]{appendix}
\usepackage[normalem]{ulem}
\usepackage{amsthm}
\usepackage{tikz}
\usepackage[labelformat=simple]{subcaption}

\definecolor{myblue}{rgb}{.93, .93, 1}

\definecolor{darkgreen}{rgb}{0,0.7,0}

\newcommand{\beq}{\begin{equation}}
\newcommand{\eeq}{\end{equation}}

\newcommand{\bpm}{\begin{pmatrix}}
\newcommand{\epm}{\end{pmatrix}}
\newcommand{\bmm}{\begin{matrix}}
\newcommand{\emm}{\end{matrix}}

\newcommand{\mU}{\mathcal{U}}

\newtheorem{theorem}{Theorem}
\newtheorem{corollary}{Corollary}
\newtheorem{lemma}{Lemma}

\graphicspath{{./Figures/}}

\begin{document}

\author{Hao Chen}
\email{chen.hao@princeton.edu}
\affiliation{
Department of Physics,
Princeton University, NJ 08544, USA
}
\affiliation{
Department of Electrical and Computer Engineering,
Princeton University, NJ 08544, USA
}

\author{Yu-Min Hu}
\affiliation{Institute for Advanced Study, Tsinghua University, Beijing, 100084, China}

\author{Wucheng Zhang}
\affiliation{
Department of Physics,
Princeton University, NJ 08544, USA
}

\author{Michael Alexander Kurniawan}
\affiliation{
Department of Physics, Hong Kong University of Science and Technology, Clear Water Bay Road, Kowloon, Hong Kong
}

\author{Yuelin Shao}
\affiliation{Beijing National Laboratory for Condensed Matter Physics and Institute of Physics,
Chinese Academy of Sciences, Beijing 100190, China}
\affiliation{School of Physical Sciences, University of Chinese Academy of Sciences, Beijing 100049, China}

\author{Xueqi Chen}
\affiliation{Department of Physics, Stanford University, Stanford, CA 94305, USA}

\author{Abhinav Prem}
\affiliation{School of Natural Sciences, Institute for Advanced Study, Princeton, New Jersey 08540, USA}

\author{Xi Dai}\thanks{corresponding author}
\email{daix@ust.hk}
\affiliation{
Department of Physics, Hong Kong University of Science and Technology, Clear Water Bay Road, Kowloon, Hong Kong
}

\date{\today}

\title{Periodically Driven Open Quantum Systems: Spectral Properties and Non-Equilibrium Steady States}
\date{\today}

\begin{abstract}
In this article, we investigate periodically driven open quantum systems within the framework of Floquet-Lindblad master equations. Specifically, we discuss Lindblad master equations in the presence of a coherent, time-periodic driving and establish their general spectral features. We also clarify the notions of transient and non-decaying solutions from this spectral perspective, and then prove that any physical system described by a Floquet-Lindblad equation must have at least one \textit{physical} non-equilibrium steady state (NESS), corresponding to an eigenoperator of the Floquet-Lindblad evolution superoperator $\mathcal{U}_F$ with unit eigenvalue. Since the Floquet-Lindblad formalism encapsulates the entire information regarding the NESS, it in principle enables us to obtain non-linear effects to all orders at once. The Floquet-Lindblad formalism thus provides a powerful tool for studying driven-dissipative solid-state systems, which we illustrate by deriving the nonlinear optical response of a simple two-band model of an insulating solid and comparing it with prior results established through Keldysh techniques. 
\end{abstract}
\maketitle


\section{Introduction}
\label{sec:intro}

Open quantum many-body systems are a cornerstone of modern physics and have been a focus of intense recent research, driven partly by the advent of NISQ-era devices \cite{preskill2018quantum}. From a theoretical perspective, open systems are of fundamental interest since the interplay between coherent Hamiltonian dynamics and dissipative processes often gives rise to novel phenomena with no equilibrium counterpart (see Refs.~\cite{lidarreview,weimerreview} for a review). Moreover, since realistic systems are always coupled to an external environment, understanding the effect of such an environment on the equilibrium phases of matter presents an important challenge. Aside from certain special cases, however, it is unfeasible to exactly treat the bath dynamics and various approximations are used to capture the effects of the environment on the system whilst only keeping track of the system degrees of freedom. 

Often, it is appropriate to work in the Markovian regime, where the system is appropriately described in terms of a quantum master equation~\cite{oxford}, of which the Lindblad (or GKLS) master equation~\cite{lindblad1976,gorini1976completely} is the most widely applicable and has contributed significantly to our understanding of quantum systems coupled weakly to an environment~\cite{app1,app2,app3,app4,app5,app6,app7,app8,app9,app10,app11,app12}. In short, the Lindblad equation provides a general description of the dynamics of a system that is weakly coupled to a large memory-less reservoir (see, however, Ref.~\cite{dharlimits} for limitations of the Lindblad equation): under these standard Born-Markov assumptions, the system's density operator evolves according to~\cite{oxford}
\beq
    \frac{d\rho}{dt}=\mathcal{L}\rho = -i[H,\rho] + \sum_\mu\Gamma_\mu \left(L_\mu \rho L_\mu^\dagger - \frac{1}{2}\{L_\mu^\dagger L_\mu,\rho\}\right) \, ,\label{eq:Leq}
\eeq
(in units where $\hbar = 1$). In this equation, the \emph{Liouville superoperator} $\mathcal{L}$ is a linear operator acting on the Liouville space $\mathscr{H}\otimes\mathscr{H}^*$, where $\mathscr{H}$ denotes the Hilbert space of the system alone. The Liouville superoperator consists of two distinct parts: the first part corresponds to the Hamiltonian $H$ of the system and generates the coherent evolution, while the second part contains so-called ``jump operators" $L_\mu$ that capture the average influence of the system-reservoir coupling in the weakly-coupled Markovian limit. The coupling strengths of these jump operators are specified by nonnegative damping rates $\Gamma_\mu\geq 0$.

The most important property of the Lindblad equation is that it generates a quantum dynamical map that is \emph{completely positive and trace-preserving} (CPTP)~\cite{lindblad1976,gorini1976completely,AJP}, which implies that the time evolution generated by $\mathcal{L} $ will keep ${\rm Tr}\rho(t)$ unchanged and will preserve the semi-definiteness of $\rho(t)$ at any future time $t$ \footnote{The condition of the dynamics being ``completely'' positive is more general than simply keeping a non-negative $\rho(t_0)$ non-negative at future times $t>t_0$, which is the positivity condition. In other words, not all positive dynamics are completely positive: see Ref.~\cite{AJP} for more details.}. The CPTP conditions ensure that a physical initial state $\rho(t_0)$ remains physical at all future times $t>t_0$, where by a physical (mixed) state we mean a positive semi-definite and Hermitian density matrix with unit trace. Note that the dynamical map generated by $\mathcal{L}$ is CPTP for any given choice of Hamiltonian $H$ and jump operators $L_\mu$ as long as $\Gamma_\mu\ge 0$. Specifically, if all $\Gamma_\mu=0$, the Lindblad equation reduces to the von Neumann equation $\dot{\rho} = -i[H,\rho]$ that describes the unitary evolution of an isolated quantum system.

In the presence of a time-independent Hamiltonian and time-independent jump operators $L_\mu$, the evolution from a generic initial state typically contains two components: a transient component that eventually decays out, and a non-decaying component that is a linear combination of eigenoperators of $\mathcal{L}$ whose eigenvalues have zero real parts (see Appendix~\ref{app:tindp}). As a consequence of the relation between the dynamics of the system density matrix $\rho(t)$ and the eigenoperators of the superoperator $\mathcal{L}$, the spectrum of $\mathcal{L}$ has recently been an object of significant interest~\cite{spectral_2000,spectrum_Jiang,spectrum_proof,spectrum_Floquet,Gary_contact}. It is worth noting that each individual eigenoperator of $\mathcal{L}$ is generically \textit{not} a physical density matrix as most of them are non-Hermitian and traceless.

Recently, attention has turned towards the dynamics of periodically driven open quantum systems, where the competition between coherent external driving and incoherent dissipation has been experimentally observed to stabilize non-trivial non-equilibrium steady states (NESS) in e.g., coupled arrays of circuit QED resonators~\cite{fitzpatrick2017,fedorov2021}. In such open Floquet systems~\cite{open_floquet_review,open_floquet1,open_floquet2,open_floquet3,is_there}, the Liouville superoperator satisfies $\mathcal{L}(t+T) = \mathcal{L}(t)$ (with $T$ the period). For closed quantum systems, such time-dependent problems are solved using Floquet theory, which generalizes the energy eigenstates of time-independent systems to Floquet eigenmodes with the same period $T$ (see Ref.~\cite{floquet_notes} and Appendix~\ref{app:floquet} for details). However, due to the intrinsic non-unitary evolution of open quantum systems, strategies from the Floquet theory for closed systems cannot be directly applied in this setting~\cite{open_floquet_nonmarkovian}. For instance, prior studies have shown that the concept of a Floquet Hamiltonian (or more generically, a Floquet Liouville superoperator), which exists for periodically driven closed quantum systems, does not always exist for open quantum systems~\cite{is_there,floquet_lindbladian1,floquet_lindbladian2,floquet_lindbladian3} and the non-unitary evolution can lead to difficulties in properly defining Floquet eigenmodes (see Ref. \cite{no_floquet_contact} for an algorithm for getting NESS without involving Floquet formalism). 

In this article, we formulate the Floquet theory of periodically driven open quantum systems which are described by time-dependent Lindblad master equations. We define the Floquet-Lindblad eigenmodes and prove the spectral properties of the Floquet-Lindblad evolution superoperator $\mU_F$ (defined in~Eq.\eqref{eq:UF}). As a consequence of these spectral features, we show that, as in the time-independent situation, there always exists at least one physical eigenoperator $\rho_0$ of $\mU_F$ with unit eigenvalue, which corresponds to the time-dependent non-equilibrium steady state (NESS) of the system. To illustrate this formalism, we consider a simple two-band model of an insulating solid and derive its nonlinear optical response, which was previously established using Keldysh Greens' function techniques in Ref.~\cite{nagaosa_2016}.

The rest of the paper is organized as follows: in Sec.~\ref{sec:eigen}, we introduce the setup of Floquet-Lindblad equations, define the Floquet-Lindblad evolution superoperator $\mU_F$ and the corresponding time-periodic Floquet-Lindblad eigenmodes, and explain the physical importance of the spectrum of $\mU_F$. Then, the spectral properties of $\mU_F$ are discussed in Sec.~\ref{sec:spectral}, particularly for the situation that $\mU_F$ is not diagonalizable. In Sec.~\ref{sec:optical}, we illustrate the Floquet-Lindblad formalism by deriving the nonlinear optical response in a simple two-band insulator within certain approximations. We conclude in Sec.~\ref{sec:cncls} with a brief discussion of future directions.


\section{Floquet-Lindblad eigenmodes}
\label{sec:eigen}
We focus on the problem of how an open quantum system interacts with a periodically varying time-dependent external field with period $T$. We limit ourselves to the simple situation in which the external field is only coupled to the system, such that the environment remains at thermal equilibrium at all times. Moreover, we assume throughout that we are in a regime in which the  Born-Markov approximations hold. Consequently, after integrating out the degrees of freedom of the environment, the equation of motion for the reduced density matrix $\rho(t)$ of the system takes the form Eq.~\eqref{eq:Leq}, i.e., for each fixed time $t$ the dissipation terms are in Lindblad form, with both the Hamiltonian $H(t)$ and the jump operators $L_\mu(t)$ depending explicitly on time $t$ and having the same period $T$ as the external field~\cite{oxford}. As the microscopic derivation of the equation of motion in Lindlbad form involves the Floquet theory, we will refer to it as the \emph{Floquet-Lindblad equation}~\cite{Fl1,FL2,FL3,open_floquet3,FLSSgeneral,timecrystal,floquet_lindbladian3} throughout this paper. 

The Floquet-Lindblad equation takes the form:
\beq\label{eq:tdeplind}
\begin{aligned}
    \frac{d\rho}{dt} &= {\mathcal{L}}(t)\rho \\ &\equiv -i[H(t),\rho] + \sum_\mu\Gamma_\mu \left(L_\mu(t) \rho L_\mu^\dagger(t) - \frac{1}{2}\{L_\mu^\dagger(t) L_\mu(t),\rho\}\right),
\end{aligned}
\eeq
where the Liouville superoperator is time-periodic $\mathcal{L}(t) = \mathcal{L}(t+T)$. In this article, we only consider systems with a finite-dimensional Hilbert space $\mathscr{H}$, with the corresponding Liouville space given by $\mathscr{H}\otimes\mathscr{H}^*$. We emphasize here that the Floquet-Lindblad equation only applies to the situations where the Born-Markov approximation holds \cite{oxford}, otherwise the dynamics of the environment will generically make the system's dynamics non-local in time and thus cannot be simply expressed by a superoperator $\mathcal{L}(t)$. In more general cases beyond Born-Markov approximation, not only does the master equation fail to exhibit the structure described in Eq. \eqref{eq:tdeplind}, but also a time-periodic external driving does not necessarily result in periodic time dependence. 

Now, we adopt the Floquet theory to solve Eq. \eqref{eq:tdeplind}. First, we focus on the time-evolution superoperator that describes the evolution from an initial time $t_0$ to a future time $t$:
\beq
\mathcal{U}(t,t_0) = \mathcal{T}e^{\int_{t_0}^t\mathcal{L}(t')dt'} \, .
\eeq
 Since $\mathcal{L}(t)$ is periodic with period $T$, we can split the entire time interval from $t_0$ to $t$ as $t-t_0 = mT+\tau$, where $m\in\mathbb{N}_0$ and the residual time $\tau\in[0,T)$. Then,
\beq
\mathcal{U}(t,t_0) = \mathcal{T} e^{\int_{t_0}^{t_0+\tau}\mathcal{L}(t')dt'}(\mathcal{U}_F)^m = \mathcal{U}(t_0+\tau,t_0)(\mathcal{U}_F)^m,\label{eq:UF}
\eeq
where the \emph{Floquet-Lindblad evolution superoperator} is defined as $\mathcal{U}_F\equiv\mathcal{U}(t_0+T,t_0)$ and describes the stroboscopic evolution of the system. Next, we define the eigenoperators $\rho_j$ of $\mU_F$ and denote the corresponding eigenvalue as $e^{\lambda_jT}$:
\beq
    \mU_F\rho_j = e^{\lambda_jT}\rho_j,
\eeq
resembling its counterpart $U_F\ket{u_j} = e^{-i\epsilon_jT}\ket{u_j}$ in non-dissipative Floquet systems. Similarly to the quasi-energy $\epsilon_j$ in that case, the imaginary part of the $\lambda_j$ is determined only modulo $2\pi/T$. It is worth emphasizing that these eigenoperators $\rho_j$ are not guaranteed to be physical density operators, and that all the eigenoperators may not form a complete basis of the Liouville space because $\mU_F$ is not always diagonalizable. 

It is straightforward to show (see Appendix~\ref{app:evol_floquet_modes}) that the time evolution of $\rho_j(t)$ can be written as a periodic function $\varrho_j(t)$ modulated by an envelope function:
\beq\label{eq:evol}
    \rho_j(t)\equiv\mU(t,t_0)\rho_j = e^{\lambda_j(mT+\tau)}\varrho_j(t_0+\tau), 
\eeq
where $\varrho_j(t_0+\tau)\equiv \mU(t_0+\tau,t_0)e^{-\lambda_j\tau}\rho_j$ is constructed such that that $\varrho_j(t+T)=\varrho_j(t)$ and is called a \emph{Floquet-Lindblad eigenmode}. Thus, the physical significance of the eigenoperator $\rho_j$ is that, if the system begins to evolve from an eigenoperator $\rho_j=\varrho_j(t_0)$ at time $t_0$, it returns to itself up to a factor $e^{\lambda_j T}$ after one period. Amongst all the eigenoperators, those with eigenvalues $|e^{\lambda_jT}| = 1$ are special as these at most accumulate a \textit{phase factor} at the end of each period i.e., these are non-decaying eigenmodes and are referred to as ``coherent Floquet modes". Moreover, the eigenoperator(s) $\rho_0$ with eigenvalue $e^{\lambda_0T}=1$ evolves with period $T$ because $\lambda_0=0$ implies $\rho_0(t) = \varrho_0(t)$ by Eq. \eqref{eq:evol}. Such periodically evolving states with period $T$ are referred to as \emph{Floquet-Lindblad steady states} \cite{FLSSgeneral, timecrystal,floquet_lindbladian3}. If such an eigenoperator $\rho_0$ is physical, its evolution $\rho_0(t)$ is a NESS of the system. In general, the NESS for the Floquet-Lindblad equation is defined as any physical density operator $\tilde{\rho}(t)$ that solves the equation and obeys $\mathcal{U}_F \tilde{\rho}(t_0) = \tilde{\rho}(t_0)$. 

In the next Section, we list and prove the spectral properties of the Floquet-Lindblad evolution superoperator $\mU_F$, showing that:
\begin{enumerate}
    \item For any Floquet-Lindblad equation, there exists a subspace of the Liouville space spanned by eigenoperators $\rho_j$ of $\mU_F$ with eigenvalues $|e^{\lambda_jT}| = 1$. Generic operators $\rho(t_0)$ initialized in this \emph{non-decaying subspace} do not decay during their time-evolution, while other generic operators initialized outside the non-decaying subspace will eventually decay out. Moreover, the non-decaying subspace depends on the choice $t_0$ in $\mU_F=\mU(t_0+T,t_0)$, indicating that a non-decaying solution $\rho(t)$ is only guaranteed to remain within the non-decaying subspace at the end of periods i.e., when $t = t_0+mT$.
    
    \item $\mU_F$ always has a unit eigenvalue $e^{\lambda_0T}=1$, whose corresponding eigenoperators $\rho_0^{(1)},...,\rho_0^{(n)}$ are Floquet-Lindblad steady states. The space spanned by them, called the \emph{steady space}, is a subspace of the non-decaying subspace and depends on the choice $t_0$ as well. Moreover, there is always at least one eigenoperator in the steady space that is physical, corresponding to the NESS of the system. 
    \item Given any physical initial state $\rho(t_0)$, it always has non-vanishing support on the non-decaying subspace i.e., it must have a non-decaying component which lives in the non-decaying subspace. The time-evolution of this non-decaying component is a linear combination of time-evolved eigenoperators $\rho_j(t)$ with eigenvalues of $|e^{\lambda_jT}| = 1$, which always include a physical NESS $\rho_0(t)$.
    
    It is important to distinguish between the non-decaying component of an initial state and that of a time-dependent solution. The former is a linear combination of $\rho_j$ with eigenvalues of $|e^{\lambda_jT}| = 1$ and thus lies in the non-decaying subspace, while the latter is a linear combination of the evolved eigenoperators $\rho_j(t)$ or eigenmodes $\varrho_j(t)$, which is not guaranteed to be in the non-decaying subspace unless at the end of each period. The properties of the non-decaying component of an initial state were previously discussed from the perspective of Floquet dynamical symmetries (FDS) in Ref.~\cite{timecrystal}.
    
    \item In the special case that $e^{\lambda_0T} = 1$ is non-degenerate, the corresponding eigenoperator $\rho_0$ is guaranteed to be physical and its time evolution $\rho_0(t)$ is the unique physical NESS of the system. In other words, this $\rho_0$ is a component of the non-decaying part of any physical initial state.
    
    \item If the non-decaying part of an initial state $\rho(t_0)$ is entirely in the steady space, it shows a periodic evolution with period $T$ after the transient decays out, i.e. it will gradually converge to a NESS. If the non-decaying part of $\rho(t_0)$ is not entirely in the steady space, it may evolve with a period longer than $T$ after the transient part decays out, which means it does not converge to a NESS by definition. 
\end{enumerate}


\section{Properties of Floquet-Lindblad eigen solutions}
\label{sec:spectral}

In this section, we present and mathematically prove the properties of the spectrum $\{(e^{\lambda_jT},\rho_j)\}$ of the Floquet-Lindblad evolution operator $\mU_F$. We only assume that the Liouville superoperator is time-periodic $\mathcal{L}(t) = \mathcal{L}(t+T)$ and takes the Lindblad form (see Eq. \eqref{eq:tdeplind}),
where the positive damping rates $\Gamma_\mu>0$ ensures the CPTP nature of the dynamical map. The spectral properties of the Floquet-Lindblad equations are similar to those of the time-independent Lindblad equations~\cite{spectrum_proof}, which are reviewed in Appendix~\ref{app:tindp}. 
\begin{figure}
    \centering
    \includegraphics[]{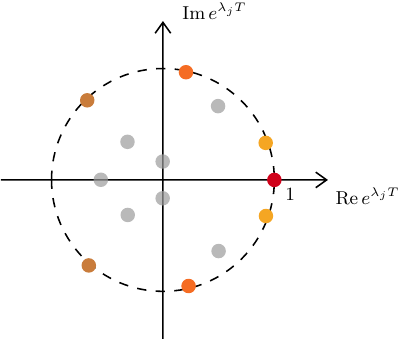}
    \caption{Eigenvalues $e^{\lambda_jT}$ of $\mU_F$ shown in the complex plane. The colored dots represent eigenvalues with norms of $1$, corresponding to the non-decaying subspace. The red dot represents the eigenvalue(s) $e^{\lambda_0T}=1$, which always exists and corresponds to the steady space. The gray dots are eigenvalues with norms smaller than $1$, contributing to transient solutions. All eigenvalues appear as conjugate pairs.}
    \label{fig:eigenval}
\end{figure}

The first three properties are closely related to the distribution of eigenvalues $e^{\lambda_jT}$, schematically illustrated in Fig.~\ref{fig:eigenval}. 
\begin{theorem}\label{theorem1}
    If $(e^{\lambda_jT},\rho_j)$ is an eigensolution, so is $(e^{\lambda_j^*T},\rho_j^\dag)$.
\end{theorem}
\begin{proof}
    This comes from the fact that
    \beq
        e^{\lambda_j^*T}\rho_j^\dag =(e^{\lambda_jT}\rho_j)^\dag = (\mU_F\rho_j)^\dag =\mU_F\rho_j^\dag.
    \eeq
    To prove the last equality, notice that $\mU_F$ is a product of piecewise-constant evolution with small enough time interval $\Delta t$: $\mU_F = \prod_{s=1}^{T/\Delta t}(I+\mathcal{L}(t_0+s\Delta t)\Delta t)$. By the definition of $\mathcal{L}$, we know $(\mathcal{L}\rho_j)^\dag = \mathcal{L}\rho_j^\dag$, implying $(\mU_F\rho_j)^\dag =\mU_F\rho_j^\dag$.
\end{proof}

\begin{corollary}\label{corrolary2.1}
    If the eigenvalue $e^{\lambda_jT}$ is real and non-degenerate, then $\rho_j = \rho_j^\dag$. Likewise, if $e^{\lambda_jT}$ is not real, then $\rho_j\neq\rho_j^\dag$. Inversely, if $\rho_j = \rho_j^\dag$, then $e^{\lambda_jT}$ is real.
\end{corollary}

\begin{theorem}\label{theorem2}
    If $e^{\lambda_jT}\neq 1$, then ${\rm Tr}\rho_j=0$.
\end{theorem}
\begin{proof}
    Due to the trace-preserving property, $e^{\lambda_jT}{\rm Tr}\rho_j = {\rm Tr}(\mU_F\rho_j) = {\rm Tr}\rho_j$. Thus ${\rm Tr}\rho_j = 0$.
\end{proof}

\begin{theorem}\label{theorem3}
    For any eigenvalue $e^{\lambda_j T}$, its modulus $|e^{\lambda_j T}|\leq 1$.    
\end{theorem}
\begin{proof}
    We prove this by contradiction and construction. Assume there exists an eigenvalue $|e^{\lambda_j T}|> 1$ and its corresponding eigenoperator is $\rho_j$. We can construct a Hermitian and positive semi-definite density operator as the initial state of the evolution:
    \beq
        \rho(t_0) = I + \varepsilon(\rho_j + \rho_j^\dag),
    \eeq
    where $\varepsilon\ll 1$ is a sufficiently small real number so that $\rho(t_0)$ remains positive semi-definite. We denote ${\rm Tr}I=D$.
    Here we assume $\rho_j + \rho_j^\dag\neq 0$. If it is zero, we then replace it with $i(\rho_j - \rho_j^\dag)$. Now, we evolve $\rho(t_0)$ for $m\gg 1$ periods:
    \beq
        \rho(t_0+mT) = \mU_F^mI + \varepsilon(e^{\lambda_jmT}\rho_j + e^{\lambda_j^*mT}\rho_j^\dag).
    \eeq
    Since the evolution is completely positive and $I$ is positive semi-definite, $\mU_F^m I$ is also positive semi-definite, i.e., for any normalized $\ket{\psi}\in\mathscr{H}$, $\bra{\psi}\mU_F^m I\ket{\psi}$ is non-negative. Also, by the trace-preserving property, we know that ${\rm Tr}\left[\mU_F^mI\right]=D$ and then $\bra{\psi}\mU_F^m I\ket{\psi}$ is bounded by $D$.  

    As $|e^{\lambda_j T}|> 1$, we can find an integer $m$ large enough and a vector $|\psi\rangle$ such that 
    \begin{equation*}
        \left|\bra{\psi}\varepsilon(e^{\lambda_jmT}\rho_j + e^{\lambda_j^*mT}\rho_j^\dag)\ket{\psi}\right|> D.
    \end{equation*}
    Then $\bra{\psi} \rho(t_0+mT)\ket{\psi} > D$ or $\bra{\psi} \rho(t_0+mT)\ket{\psi} < 0$, which contradicts that $\rho(t_0+mT)$ must remain positive semi-definite under the evolution~\cite{lindblad1976,gorini1976completely,AJP}.
\end{proof}

Now, in the simple situation where $\mU_F$ is diagonalizable, we readily see that any state in the Liouville space is a linear combination of eigenoperators $\rho_j$ since the latter form a complete basis. Thus, an arbitrary state at initial time $\rho(t_0) = \sum_j a_j\rho_j$ evolving to time $t = t_0+mT+\tau$ becomes
\beq
    \rho(t) = \sum_j a_j\rho_j(t) = \sum_j a_je^{\lambda_j(t-t_0)}\varrho_j(t),
\eeq
where $\varrho_j(t) = \varrho_j(t_0+\tau)$ are the time-periodic Floquet-Lindblad eigenmodes. It is apparent that at a large enough time $t$, all the components with $|e^{\lambda_jT}|<1$ will eventually decay out, leaving only the non-decaying components with eigenvalues $|e^{\lambda_jT}|=1$. This aligns with our physical intuition that a solution of a Floquet-Lindblad equation usually oscillates while decaying, and may leave a non-decaying component after the transient part dies out. However, as the evolution is not unitary, the diagonalizability of $\mU_F$ is not guaranteed. We need further investigation of the spectral properties to get a comprehensive understanding of the driven-dissipative dynamics.

Therefore, we now focus on the cases where $\mU_F$ cannot be completely diagonalized. In other words, there exists at least one deficient eigenvalue $e^{\lambda_jT}$ whose geometric multiplicity $k$ is smaller than its algebraic multiplicity (i.e., degeneracy) $n$, which is also known as an exceptional point. Each deficient eigenvalue with degeneracy $n$ is associated with an $n$-dimensional subspace $V_{\lambda_j}$ of the Liouville space spanned by its eigenoperators and some other operators, such that $\mU_FV_{\lambda_j}\subseteq V_{\lambda_j}$, leading to the definition: $\mU_{F\lambda_j}\equiv\mU_F|_{V_{\lambda_j}}$ i.e, the Floquet evolution operator acting on $V_{\lambda_j}$. Then, one can properly choose the basis of $V_{\lambda_j}$, including the $k$ eigenoperators, to form a transition superoperator $\mathcal{P}_{\lambda_j} = 
    \bpm
            \rho_j^{(1)}& ... & \rho_j^{(k)} &p_j^{(1)}&...&p_j^{(n-k)}
    \epm
$, such that $\mathcal{P}^{-1}_{\lambda_j}\mU_{F\lambda_j}\mathcal{P}_{\lambda_j} = \mathcal{J}_{F\lambda_j}$ and $\mathcal{J}_{F\lambda_j}$ is the Jordan normal form. In Theorem \ref{theorem2} we proved that $\rho_j^{(1)},...,\rho_j^{(k)}$ are traceless if $e^{\lambda_jT}\neq 1$. Now we want to show that $p_j^{(1)},...,p_j^{(n-k)}$ are also traceless.

\begin{theorem}\label{theorem4}
    For every deficient eigenvalue $e^{\lambda_jT}\neq 1$, all the operators $\rho_j^{(1)},...,\rho_j^{(k)} ,p_j^{(1)},...,p_j^{(n-k)}$ spanning the transition superoperator $\mathcal{P}_{\lambda_j}$ are traceless.
\end{theorem}
\begin{proof}
    Since there are $k-1$ eigenoperators that only contribute to a diagonal block of $\mathcal{J}_{F\lambda_j}$, we can separate them from the basis spanning $V_{\lambda_j}$ and we only need to prove the case $k=1$. Supposing that $\lambda_j$ is $n$-fold degenerate, we can represent $\mU_{F\lambda_j}$ in matrix form:
    \beq
    \begin{aligned}
        &\mathcal{P}_{\lambda_j}^{-1}\mU_{F\lambda_j}\mathcal{P}_{\lambda_j} = 
        \bpm
            e^{\lambda_jT}\ \ &1&\ &\ \\
            0\ \ &e^{\lambda_jT}&1&\ \\
            \ &0&\ddots&\ddots\\
            \ &\ &\ddots&\ddots&1\\
            \ &\ &\ &0&e^{\lambda_jT}
        \epm,\\
        &\mathcal{P}_{\lambda_j} = 
        \bpm
            \rho_j^{(1)}&p_j^{(1)}&...&p_j^{(n-1)}
        \epm.
    \end{aligned}
    \eeq
    Theorem \ref{theorem2} tells us ${\rm Tr}\rho_j^{(1)}=0$; then we can use the trace-preserving property to show that
    \beq
        {\rm Tr}(\mU_{F\lambda_j}p_j^{(1)}) = {\rm Tr}\rho_j^{(1)} + e^{\lambda_jT} {\rm Tr}p_j^{(1)} = {\rm Tr}p_j^{(1)} \Longrightarrow {\rm Tr}p_j^{(1)}=0.
    \eeq
    Likewise, ${\rm Tr}p_j^{(2)} = ... = {\rm Tr}p_j^{(n-1)}=0$ can be proved by continuously applying induction.
\end{proof}

\begin{theorem}\label{theorem5}
    $\mU_F$ must have at least one eigenoperator $\rho_0$ such that $e^{\lambda_0T}=1$ while ${\rm Tr}\rho_0\neq 0$. This $\rho_0$, by definition, is a Floquet-Lindblad steady state of the system such that $\mU_F\rho_0 = \rho_0$.
\end{theorem}
\begin{proof}
    If $\mathcal{U}_F$ is assumed to be diagonalizable, the proof is trivial. This is because when all the eigenoperators form a complete basis of the Liouville space $\mathscr{H}\otimes\mathscr{H}^*$, there is at least one eigenoperator that is traceful. Given that Theorem~\ref{theorem2} shows that eigenstates with nonunit eigenvalues are traceless, we conclude that the traceful eigenoperator has a unit eigenvalue.
    
    Furthermore, without assuming $\mU_F$ is diagonalizable, we can also show the existence of a unit eigenvalue and a corresponding traceful eigenoperator. As mentioned in the proof of Theorem~\ref{theorem4}, in the situation where $\mU_F$ is not diagonalizable, the operators spanning all the transition superoperators $\mathcal{P}_{\lambda_j}$'s still form a complete basis of the Liouville space. Therefore, there must be at least one of them that is traceful. Since Theorem~\ref{theorem4} already shows that all the operators spanning $V_{\lambda_j}$ with $e^{\lambda_jT}\neq 1$ are traceless and there must be at least one traceful operator in the Liouville space, there is at least one operator in $V_{\lambda_0}$ subspace such that $e^{\lambda_0T}=1$. As a consequence, we prove the existence of $e^{\lambda_0T}=1$, see Fig.~\ref{fig:eigenval}. 
    
    We are now going to prove that at least one eigenoperator of $e^{\lambda_0T}=1$ is traceful. There are two cases:

    1. If $\mU_{F\lambda_0}$ is diagonalizable in the $e^{\lambda_0T}=1$ subspace $V_{\lambda_0}$, then all the operators spanning $V_{\lambda_0}$ are eigenoperators of eigenvalue $e^{\lambda_0T}=1$. Therefore, at least one of them has non-zero trace.

    2. The other possibility is that $\mU_{F\lambda_0}$ is not diagonalizable. However, we will show in Theorem~\ref{theorem6} that $\mU_{F\lambda_0}$ is always diagonalizable.
\end{proof}

\begin{theorem}\label{theorem6}
    If $e^{\lambda_0T}=1$ has degeneracy $n$, then there exist $n$ independent corresponding eigenoperators i.e., $\mU_{F\lambda_0}$ is diagonalizable.
\end{theorem}
\begin{proof}
    We will prove this by contradiction, extending the logic in Ref.~\cite{spectrum_proof} that proved a similar property for time-independent $\mathcal{L}$. Supposing that $\mU_{F\lambda_0}$ is not diagonalizable, we can always put the non-diagonalizable part into Jordan normal form:
    \beq
    \begin{aligned}
    \mathcal{P}_{\lambda_0}^{-1}\mU_{F\lambda_0}\mathcal{P}_{\lambda_0} &= 
        \bpm
            e^{\lambda_0T}\ \ &1&\ &\ \\
            0\ \ &e^{\lambda_0T}&1&\ \\
            \ &0&\ddots&\ddots\\
            \ &\ &\ddots&\ddots&1\\
            \ &\ &\ &0&e^{\lambda_0T}
        \epm
        \\
    &= (e^{\lambda_0T} + \mathcal{N}),
    \end{aligned}
    \eeq
    where $\mathcal{N}$ is the upper triangle part of the Jordan normal form, and is a nilpotent matrix with $\mathcal{N}^{n-1}=0$.
    Imagine that we evolve the system from $t_0$ to $t_0+mT$ $(m>n)$ by applying $\mU_F$ by $m$ times; then, in the $V_{\lambda_0}$ subspace we have:
    \beq
    \begin{aligned}
        \mU_{F\lambda_0}^m &= \mathcal{P}_{\lambda_0} (e^{\lambda_0T}+\mathcal{N})^m\mathcal{P}_{\lambda_0}^{-1}\\
        &= \mathcal{P}_{\lambda_0}\sum_{k=0}^m C_m^k e^{(m-k)\lambda_0 T}\mathcal{N}^k \mathcal{P}_{\lambda_0}^{-1}\\
        &= q^m\mathcal{P}_{\lambda_0} 
        \bpm
           1\ \ &\frac{m}{q}&\frac{m(m-1)}{2q^2}&...&\frac{C_m^{n-1}}{q^{n-1}}&\frac{C_m^n}{q^n}\\
            0\ \ &1&\frac{m}{q}&...&\frac{C_m^{n-2}}{q^{n-2}}&\frac{C_m^{n-1}}{q^{n-1}}\\
            \ &0&\ddots&\ddots&\vdots&\vdots\\
            &&\ddots&\ddots&\vdots&\vdots\\
            \ &\ & &\ddots&1&\frac{m}{q}\\
            \ &\ &\ &\ &0&1
        \epm
        \mathcal{P}_{\lambda_0}^{-1},
    \end{aligned}
    \eeq
    where $C_m^k=\bpm m\\k\epm$ is the binomial coefficient and we have used $q=e^{\lambda_0T}$ for simplicity.
    Since $q=1$, the above expression will clearly cause the dynamics to diverge in large $m$, showing a contradiction.
\end{proof}

We can relax the condition $e^{\lambda T}=1$ to $|e^{\lambda  T}| = 1$ and still show $\mU_{F\lambda}$ is diagonalizable. The proof is similar.
\begin{theorem}\label{theorem7}
    If  $|e^{\lambda_jT}| = 1$, $\mU_{F\lambda_j}$ is diagonalizable.
\end{theorem}

 In summary, despite the non-diagonalizability of $\mU_F$, we still have a complete basis of the Liouville space consisting of eigenoperators $\rho_j$'s and non-eigen operators $p_j$'s that correspond to each eigenvalue $e^{\lambda_jT}$. For any $|e^{\lambda_jT}|<1$, by noticing that for $1< s\leq n-k$, $\mU_F p_j^{(s)} = p_j^{(s-1)} + e^{\lambda_jT}p_j^{(s)}$ and $\mU_F p_j^{(1)} =  \rho_j^{(k)} + e^{\lambda_jT}p_j^{(1)}$, we realize that a linear combination of $\rho_j$'s and $p_j$'s will gradually decay out at a large time $t$, representing a transient solution.

Theorems~\ref{theorem6} and~\ref{theorem7} ensure that the subspaces associated with $|e^{\lambda_jT}| = 1$ are spanned by eigenoperators only, which directly leads to the fact that the non-decaying component of any initial state is a linear combination of eigenoperators of $\mU_F$. We will call the space spanned by all these eigenoperators as the \emph{non-decaying subspace}. 

Theorems~\ref{theorem4} and~\ref{theorem5} show that all the basis operators are traceless except for the eigenoperators with eigenvalues that equal $1$. Therefore, for any physical initial state, it has to include a traceful eigenoperator $\rho_0$ in order to be traceful itself, and thus a physical initial state must have some non-decaying component.

Next, we are going to prove that there is at least one physical eigenoperator in $V_{\lambda_0=0}$. In the situation where $e^{\lambda_0T}=1$ is non-degenerate, the corresponding unique eigenoperator $\rho_0$ gives the physical NESS $\rho_0(t)$ for any physical initial states.

\begin{theorem}\label{theorem8}
    If $e^{\lambda_0T}=1$ is non-degenerate, the corresponding $\rho_0$ is Hermitian, semi-definite, and traceful, indicating that $\rho_0$ is physical after normalizing to ${\rm Tr}\rho_0=1$.
\end{theorem}
\begin{proof}
    Theorems~\ref{theorem2} and~\ref{theorem5} already prove that $\rho_0$ is Hermitian and traceful. Thus, we only need to prove its semi-definiteness here, or equivalently, prove its positive semi-definiteness after normalizing to ${\rm Tr}\rho_0=1$. 

    We prove this by construction: Take a physical density operator as the initial state, and let it evolve in time. Since the operators spanning $\mathcal{P}$ form a complete basis, the initial state can be expressed as a linear combination of them. 
    At late times $t$, all the terms initially in the $|e^{\lambda_j T}|< 1$ subspaces decay out due to the $e^{\lambda_jmT}$ factor in Eq. \eqref{eq:evol}, leaving only the terms in the $|e^{\lambda_j T}|= 1$ subspaces, i.e., the non-decaying subspace.

    Theorems~\ref{theorem6} and~\ref{theorem7} show that the non-decaying subspace is fully spanned by eigenoperators of $\mU_F$. Therefore, after evolving for a large number of periods, at $t=mT + t_0$, the kate-timephysical state (which is hermitian) can be expressed as
    \beq\label{eq:evdecomp}
        \rho(t) = a_0\rho_0 + \sum_{j:|e^{\lambda_jT}|=1,\arg e^{\lambda_jT}\neq 0}(e^{\lambda_jmT}a_j\rho_j+h.c.),
    \eeq
    where the non-zero constant $a_0\in \mathbb{R}$.
    The completely positive nature of the dynamical evolution ensures that the above remains positive semi-definite, which means that 
    \beq
        \forall \ket{\psi}\in\mathscr{H},\ \bra{\psi}\rho(t)\ket{\psi}\geq 0.
    \eeq
    Then, denoting $\bra{\psi}\rho_j\ket{\psi}=\braket{\rho_j}$, we have
    \beq
        a_0\braket{\rho_0} + \sum_{j:|e^{\lambda_jT}|=1,\arg e^{\lambda_jT}\neq 0}(e^{\lambda_jmT}a_j\braket{\rho_j}+c.c.)\geq0.
    \eeq
    The terms in the summation must take the form $\tilde{a}_j\cos(|\lambda_j|mT+\phi_j),\ \ \tilde{a}_j\in\mathbb{R}$, and because $\lambda_j$ is purely imaginary, we have 
    \beq
        a_0\braket{\rho_0} + \sum_{j:|e^{\lambda_jT}|=1,\arg e^{\lambda_jT}\neq 0}\tilde{a}_j\cos(|\lambda_j
        |mT+\phi_j)\geq0.
    \eeq
    It can be shown that the average of the time-dependent part over enough many periods is zero:
    \beq
        \frac{1}{M}\sum_{m=1}^{M}\left(\sum_{j:|e^{\lambda_jT}|=1,\arg e^{\lambda_jT}\neq 0}\tilde{a}_j\cos(|\lambda_j
        |mT+\phi_j)\right) = 0,
    \eeq
    where $M\gg 1$. This result indicates that the time-dependent part is either constantly zero or oscillating between positive and negative. Hence, $a_0\braket{\rho_0}$ should be non-negative to keep the entire equation non-negative.

    Therefore, $\bra{\psi}\rho_0\ket{\psi}$ is either non-positive or non-negative (depending on its normalization) for all $\ket{\psi}\in\mathcal{H}$. Then, $\rho_0$ is semi-definite. After normalizing to ${\rm Tr}\rho_0=1$, it is positive semi-definite.
\end{proof}

\begin{theorem}\label{theorem9}
    Regardless of the degeneracy of $e^{\lambda_0T}=1$, there exists at least one physical steady state which is an eigenoperator of $\mU_F$ with eigenvalue $e^{\lambda_0T}=1$.
\end{theorem}
\begin{proof}
    The proof is similar to that of Theorem~\ref{theorem8}, where $a_0\rho_0$ in Eq. \eqref{eq:evdecomp} is replaced by a linear combination of the degenerate eigenoperators with eigenvalue of $1$. The rest of the proof follows as before.
\end{proof}

To conclude this Section, we emphasize that a Floquet-Lindblad equation always has at least one Floquet-Lindblad steady-state solution that is physical. Moreover, we point out that even if there does not exist a ``Floquet-Liouville superoperator'' $\mathcal{L}_F$ in Lindblad form, which resembles the Floquet Hamiltonian $H_F$ in closed Floquet systems, or even if the Floquet-Lindblad evolution operator $\mU_F$ is not fully diagonalizable, $\mU_F$ is always diagonalizable in the non-decaying subspace so that the non-decaying component of any initial state $\rho(t_0)$ can be expressed as a linear combination of eigenstates $\rho_j$'s of $\mU_F$ whose eigenvalues have modulus $|e^{\lambda_jT}|=1$. We also note that starting from any physical initial state, there is always a non-decaying component in the evolution, which is a linear combination of the time-evolved eigenoperators $\rho_j(t)$ with $|e^{\lambda_jT}|=1$, and there is always a physical NESS $\rho_0(t)$ in this combination. The spectral properties of Floquet-Lindblad equations and time-independent Lindblad equations are concisely summarized in Table~\ref{table:1} for comparison. 

\begin{table*}
\begin{center}
\begin{tabular}{ l c c }
\hline\hline
 Spectral properties of & $\mathcal{L}$ of time-independent Lindblad equations & $\mU_F$ of Floquet-Lindblad equations \\ 
 \hline
 Eigenvalue equation & $\mathcal{L}\rho_j = \lambda_j\rho_j$ & $\mU_F\rho_j = e^{\lambda_jT}\rho_j$ \\  
 Eigenvalues & ${\rm Re}[\lambda_j]\leq 0$ & $|e^{\lambda_jT}|\leq 1$\\
 Conjugate pairs & If $\mathcal{L}\rho_j = \lambda_j\rho_j$, then $\mathcal{L}\rho_j^\dag = \lambda_j^*\rho_j^\dag$ & If $\mU_F\rho_j = e^{\lambda_jT}\rho_j$, then $\mU_F\rho_j^\dag = e^{\lambda_j^*T}\rho_j^\dag$\\
 Non-hermitian & $\rho_j \neq \rho_j^\dag$ as long as ${\rm Im}\lambda_j \neq 0$ & $\rho_j \neq \rho_j^\dag$ as long as ${\rm Im}e^{\lambda_jT} \neq 0$\\
 Traceless & If $\lambda_j\neq 0$, ${\rm Tr}\rho_j = 0$ & If $e^{\lambda_jT}\neq 1$, ${\rm Tr}\rho_j = 0$\\
 No complete basis & $\{\rho_j\}$ DO NOT form a complete basis & $\{\rho_j\}$ DO NOT form a complete basis\\
 Non-decaying subspace & Spanned by $\{\rho_j|{\rm Re}[\lambda_j]=0\}$ & Initial time $t_0$ dependent, spanned by $\{\rho_j||e^{\lambda_jT}| = 1\}$\\
 Physical NESS &\ \ \ \ \  $\mathcal{L}\rho_0 = 0$, $\rho_0$ is a physical density operator\ \ \ \ \  &\ \ \ \ \   $\mU_F\rho_0 = \rho_0$, $\rho_0(t)$ is a physical density operator\ \ \ \ \  \\
 \hline
\end{tabular}
\caption{Spectral properties of time-independent Lindblad equations and Floquet-Lindblad equations.}
\label{table:1}
\end{center}
\end{table*}


\section{Nonlinear Optical Effects in Solids}
\label{sec:optical}
In this Section, we apply the Floquet-Lindblad formalism to study the nonlinear optical responses of a simple insulating solid-state system described by two Bloch bands (a valence band and a conduction band), and compare the results with those from a previous study~\cite{nagaosa_2016} on a similar problem that used the Keldysh Greens' functions combined with Floquet theory~\cite{keldysh1,keldysh2,keldysh3,keldysh4}. This example illustrates that the Floquet-Lindblad method can describe coupling the system with baths at arbitrary temperatures, can go beyond the rotating-wave approximation (RWA) easily, and allows us to obtain linear effects in a non-perturbative manner. Specifically, we will investigate the electric current $J(t)$ in the insulating solid induced by the periodic driving from an external electromagnetic field. Moreover, we will focus on the current response in the steady state. 

\begin{figure}
    \centering
    \includegraphics[]{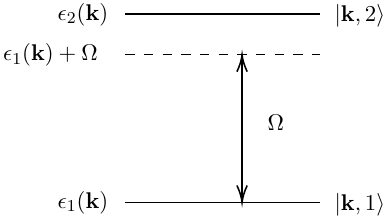}
    \caption{The structure of the effective two-level system at a quasi-momentum $\mathbf{k}$. Indices $1$ and $2$ respectively denote the valence band and the conduction band, each having its own energy dispersion $\epsilon_{1,2}^0(\mathbf{k})$. The applied time-periodic light field has angular frequency $\Omega$.}
    \label{fig:optical_trans}
\end{figure}

First, we write the Hamiltonian of the system that is driven by an external periodic drive. Among the many energy bands in an insulator, we focus on the two bands close to the Fermi surface: a fully occupied valance band and an empty conduction band that is involved in the transition induced by a monochromatic electric field (light): $E(t) = Ee^{-i\Omega t} + E^*e^{i\Omega t}$, see Fig.~\ref{fig:optical_trans}. The many-body Hamiltonian of the solid system in the absence of the driving field and within the subspace of these two energy bands is
\beq
    H^0 = \sum_\mathbf{k}\sum_{m=1}^2 \epsilon_m^0(\mathbf{k}) c_{\mathbf{k},m}^\dag c_{\mathbf{k},m},
\eeq
where $c_{\mathbf{k},m}^\dag$ denotes the creation operator of the electron at quasi-momentum $\mathbf{k}$ and energy band $m$. The band indices $m=1,2$ respectively denote the valence and conduction bands, each of which having its own energy dispersion $\epsilon_{1,2}^0(\mathbf{k})$. The coupling between the electrons and the light field introduces a periodically time-dependent term in the Hamiltonian \cite{nagaosa_2016}:
\beq\label{eq:tdeph}
    H(t) = H^0 + A(t)V^0,
\eeq
where $V^0 = \sum_{\mathbf{k},m,n} c_{\mathbf{k},m}^\dag v^0_{mn}(\mathbf{k}) c_{\mathbf{k},n}$ is the current operator \emph{in the absence of the driving field} (see Appendix \ref{app:nonlinearopt}), and $A(t) = -iAe^{-i\Omega t} + iA^*e^{i\Omega t}$ is the vector potential corresponding to the driving field: $E(t) = -\frac{\partial A(t)}{\partial t}$, such that its amplitude is $A = E/\Omega$. 

Since the form of the Hamiltonian \eqref{eq:tdeph} ensures that quasi-momentum $\mathbf{k}$ is conserved during evolution, we can treat each $\mathbf{k}$ sector as an independent two-level single-particle quantum system to be dealt with separately. In the basis spanned by single particle states $\{\ket{\mathbf{k},1}, \ket{\mathbf{k},2}\}$, the Hamiltonian of the two-level system is represented by a 2-by-2 matrix:
\beq\label{tdeph2}
    \bm{h}(\mathbf{k},t) = \bpm
        \epsilon_1^0(\mathbf{k}) + A(t)v^0_{11}(\mathbf{k})  &  A(t)v^0_{12}(\mathbf{k})\\
         A(t)v^0_{21}(\mathbf{k}) & \epsilon_2^0(\mathbf{k}) + A(t)v^0_{22}(\mathbf{k})
    \epm.
\eeq
We now apply the Shirley-Floquet formalism~\cite{shirley_floquet} (discussed in Appendix~\ref{app:SF} and~\ref{app:SFL}), which expands the time dependent Hamiltonian \eqref{tdeph2} into an infinite dimensional matrix consisting the Fourier coefficients of $\bm{h}(\mathbf{k},t)$. For simplicity of the demonstration, we will work within the RWA by only keeping a 2-by-2 block of the entire Shirley-Floquet Hamiltonian in each sector $\mathbf{k}$~\cite{nagaosa_2016} (see Appendix~\ref{app:nonlinearopt} for more details):
\beq\label{eq:SFHRWA}
\begin{aligned}
    \bm{h}_{SF}^{\rm RWA}(\mathbf{k}) &= 
    \bpm
        \epsilon_1^0(\mathbf{k}) + \Omega & -iA^*v_{12}^0(\mathbf{k})\\
        iAv_{21}^0(\mathbf{k}) & \epsilon_2^0(\mathbf{k})
    \epm
    \equiv \epsilon I + \bm{d}\cdot\bm{\sigma}\\
    &=\bpm
        \epsilon + d_z & d_x - id_y\\
        d_x + id_y & \epsilon - d_z
    \epm.
\end{aligned}
\eeq
We will use real parameters $\epsilon$ and $\bm{d} = (d_x,d_y,d_z)$ to denote them for simplicity. 

Then, we couple the system with a photonic heat bath and write down the matrix equation for solving Floquet-Lindblad eigenmodes using the Shirley-Floquet formalism discussed in Appendix~\ref{app:SFL} 
\beq\label{eq:FL}
\begin{aligned}
    \lambda_j\bm{\varrho}_j^{\rm RWA}(\mathbf{k})
    &= -i[\bm{h}_{SF}^{\rm RWA}(\mathbf{k}),\bm{\varrho}_j^{\rm RWA}(\mathbf{k})] \\
    + \sum_{\mu=1,2}\Gamma_\mu(\mathbf{k})& \left(\bm{L}_\mu \bm{\varrho}_j^{\rm RWA}(\mathbf{k}) \bm{L}_\mu^\dagger - \frac{1}{2}\{\bm{L}_\mu^\dagger \bm{L}_\mu,\bm{\varrho}_j^{\rm RWA}(\mathbf{k})\}\right),
\end{aligned}
\eeq
where $\bm{L}_1 = \sigma_+$ and $\bm{L}_2 = \sigma_-$ represent the emission/absorption of energy to/from the heat bath, respectively. At any quasi-momentum $\mathbf{k}$, the corresponding damping rates $\Gamma_{1,2}$ are controlled by the average photon number $N$ in the heat bath \cite{oxford}:
\beq
    \Gamma_1 = \gamma_0(N+1),\ \ \ \Gamma_2 = \gamma_0N,
\eeq
where $\gamma_0$ is a positive coupling constant given by the microscopic interaction between the photons in the bath and the electrons in the solid. The average number $N$ of photons at a certain energy in the bath is determined by the bath temperature via the Planck distribution \cite{oxford}:
\beq
    N = \frac{1}{\exp(\beta \varepsilon)-1},
\eeq
where $\varepsilon = \epsilon_2^0(\mathbf{k}) - \epsilon_1^0(\mathbf{k})$ is the photon energy and $\beta$ is the inverse temperature of the heat bath. As the bath temperature is encoded in $\Gamma_1$ and $\Gamma_2$, the Floquet-Lindblad method is capable of discussing the system coupled to heat baths at arbitrary temperatures. In this setup, the emission rate $\Gamma_1$ is always larger than the absorption rate $\Gamma_2$. Specifically, at zero temperature, $N=0$ so that $\Gamma_2 = 0$ and $\Gamma_1=\gamma_0$ is positive. 

Next, we solve the Floquet-Lindblad steady state by setting the eigenvalue $\lambda_j=0$. To solve Eq.~\eqref{eq:FL}, we parameterize the steady-state density matrix as $\bm{\rho}_{ss}^{\rm RWA}(\mathbf{k}) = \bm{\varrho}_{0}^{\rm RWA}(\mathbf{k}) = \frac{1}{2}(I + \braket{\bm{\sigma}}_{ss}^{\rm RWA}\cdot\bm{\sigma})$, imposing the Hermitian and unit trace conditions. Then, (suppressing the $\mathbf{k}$ index) Eq.~\eqref{eq:FL} becomes
\beq
\begin{aligned}
    0 &= 
    \bpm
        -\frac{\gamma}{2} & -2d_z & 2d_y\\
        2d_z & -\frac{\gamma}{2} & -2d_x\\
        -2d_y & 2d_x & -\gamma
    \epm
    \bpm
        \braket{\sigma_x}_{ss}^{\rm RWA}\\
        \braket{\sigma_y}_{ss}^{\rm RWA}\\
        \braket{\sigma_z}_{ss}^{\rm RWA}
    \epm
    +\bpm
        0\\
        0\\
        \gamma_0
    \epm\\
    &\equiv \bm{G}\braket{\bm{\sigma}}_{ss}^{\rm RWA} + \bm{b} \, ,
\end{aligned}
\eeq
where $\gamma = \Gamma_1+\Gamma_2$ and $\gamma_0 = \Gamma_1-\Gamma_2$. The unique Floquet-Lindblad steady state can be solved by inverting the matrix $\bm{G}$ \cite{oxford} \footnote{In the situation where there is more than one Floquet-Lindblad steady state, $\bm{G}$ becomes non-invertible and we will need to analyze its null space to obtain all the steady states.}:
\beq
    \braket{\bm{\sigma}}_{ss}^{\rm RWA} = -\bm{G}^{-1}\bm{b}=\bpm
        \frac{4(4d_xd_z + d_y\gamma)\gamma_0}{\gamma(8d_x^2 + 8d_y^2 + 16d_z^2 + \gamma^2)}\\
        -\frac{4(-4d_yd_z + d_x\gamma)\gamma_0}{\gamma(8d_x^2 + 8d_y^2 + 16d_z^2 + \gamma^2)}\\
        \frac{(16d_z^2 + \gamma^2)\gamma_0}{\gamma(8d_x^2 + 8d_y^2 + 16d_z^2 + \gamma^2)}
    \epm \, .
\eeq
With the NESS solution $\rho_{ss}(t)$, we can calculate the expectation values of any physical observable $O$ in the steady state by $\braket{O}_{ss}(t)={\rm Tr}[O\rho_{ss}(t)]$. In the present scenario, the physical quantity of interest is the current, whose second-order response to the periodic driving potential contains two parts: direct current (DC) and second harmonic generation (SHG), and their steady-state values are contributed to by every momentum sector $\mathbf{k}$. The DC part is also called the ``shift current" and the corresponding current operator in a particular momentum sector $\mathbf{k}$ can be expressed by real parameters $b_0,b_x,b_y,b_z$ as (see Appendix~\ref{app:nonlinearopt})
\beq
    \bm{v}^{\rm RWA}(\mathbf{k})\equiv b_0I + \bm{b}\cdot\bm{\sigma}
    =\bpm
        b_0 + b_z & b_x - ib_y\\
        b_x + ib_y & b_0 - b_z
    \epm.
\eeq
As a result, the steady-state value of the DC response contributed by momentum sector $\mathbf{k}$ is easily expressed by $\bm{\rho}_{ss}^{\rm RWA}(\mathbf{k})$ as
\beq
    J_{\rm DC}(\mathbf{k}) = {\rm Tr}\left(\bm{v}^{\rm RWA}(\mathbf{k})\bm{\rho}_{ss}^{\rm RWA}(\mathbf{k})\right) = b_0 + \bm{b}\cdot\braket{\bm{\sigma}}_{ss}^{\rm RWA}.
\eeq
The total DC response is a summation of contributions coming from all the momentum sectors. If we set the temperature to zero by taking $\Gamma_2 = 0$, then $\gamma = \gamma_0 = \Gamma_1$, and we find that the above result is qualitatively similar to the result obtained in Ref.~\cite{nagaosa_2016} using Keldysh Greens' functions. We attribute the quantitative difference between the two results to the nature of the bath: in Ref.~\cite{nagaosa_2016}, the system was coupled to a fermionic bath, while our coupling is to a bosonic (photonic) bath. We evaluate the SHG part of the response in Appendix~\ref{app:nonlinearopt}.

The above demonstrative calculation is in the RWA as we only keep the two lowest-frequency components in the entire Fourier series, reducing the Shirley-Floquet Hamiltonian at each momentum sector $\mathbf{k}$ to a 2-by-2 matrix \eqref{eq:SFHRWA}. To go beyond RWA in numerical calculations, one can simply include more terms with higher frequencies in the infinite dimensional Shirly-Floquet Hamiltonian matrix, getting nonlinear optical responses up to a higher order all at once.



\section{Conclusion}
\label{sec:cncls}

In this article, we have discussed the spectral features of the Floquet-Lindblad evolution superoperator $\mathcal{U}_F$ and their consequences on the resulting NESS. Further, we have illustrated how this formalism allows us to obtain nonlinear optical effects by considering a simple model of an insulating solid and contrasting our results with those obtained prior through Keldysh techniques. The Floquet-Lindblad formalism is thus well suited for describing the non-equilibrium physics of quantum many-body systems as well as that of nonlinear optical systems. In the future, it would be interesting to verify the regimes of validity of perturbative methods (such as the Magnus or van Vleck high-frequency expansions \cite{FLSSgeneral,floquet_lindbladian3}) by comparing them with the Floquet-Lindblad formalism. It would also be interesting to extend the Floquet formalism to the case of non-Markovian but still completely positive dynamics, which goes beyond the Born-Markov approximation we have considered here.


\begin{acknowledgements}
    The authors are especially grateful to Haowei Chen for enlightening and detailed discussions that helped to clarify many concepts involved in this project. We also thank Biao Lian for his support in various aspects. The work described in this paper was supported by a fellowship award from the Research Grants Council of the Hong Kong Special Administrative Region, China (Project No. HKUST SRFS2324-6S01). The authors of this paper also receive support from the National Science Foundation under award DMR-2141966 (H.C.); NSFC under Grant No. 12125405 (Y-M.H.);
    the Sivian Fund at the Institute for Advanced Study and the U.S. Department of Energy, Office of Science, Office of High Energy Physics under Award Number DE-SC0009988 (A.P.).
\end{acknowledgements}



\appendix


\section{Spectral properties of time-independent Lindbladian}
\label{app:tindp}

In this Appendix, we write down (and sketch proofs for) the properties of the eigensolutions of a time-independent Liouville superoperator. The CPTP nature of the dynamical map will play a key role in establishing several of the following lemmas. The eigenvalue equation takes the form:
\beq
\mathcal{L}\rho_j = \lambda_j\rho_j \, ,
\eeq
where $\mathcal{L}$ is the Lindbladian superoperator, $\lambda_j$ is the (generally comples) eigenvalue, and $\rho_j$ is the corresponding eigenoperator, which is not assumed to be a physical density matrix (unit trace, Hermitian, and positive definite). If we set $\rho_j$ as the initial state of the dynamics (where we set the initial time $t_0=0$ without loss of generality), the state of the system at a later time $t$ is given by 
\beq
    \rho_j(t) \equiv e^{\mathcal{L}t}\rho_j = e^{\lambda_j t}\rho_j,
\eeq
such that ${\rm Re}[\lambda_j]$ governs the decay of $\rho_j$ and ${\rm Im}[\lambda_j]$ makes it oscillate.

\begin{lemma}\label{lemma1}
    For any eigenvalue $\lambda_j$, its real part ${\rm Re}[\lambda_j]\leq 0$. (The real part of the eigenvalues is responsible for the relaxation to the steady state.)
\end{lemma}

\begin{lemma}\label{lemma2}
    If $(\lambda_j,\rho_j)$ is a pair of eigensolutions, so is $(\lambda_j^*,\rho_j^\dag)$.
\end{lemma}
\begin{proof}
    $(\mathcal{L}\rho_j)^\dag = \mathcal{L}\rho_j^\dag = (\lambda_j\rho_j)^\dag = \lambda_j^*\rho_j^\dag$
\end{proof}

\begin{lemma}\label{lemma3}
    If $\lambda_j$ is real and non-degenerate, then $\rho_j = \rho_j^\dag$. 
    Likewise, if $\lambda_j$ is not real, then $\rho_j\neq\rho_j^\dag$.
    Inversely, if $\rho_j = \rho_j^\dag$, then $\lambda_j$ is real.
\end{lemma}

\begin{lemma}\label{lemma4}
    If $\lambda_j\neq 0$, then ${\rm Tr}\rho_j=0$. 
\end{lemma}
\begin{proof}
    Due to the trace-preserving property, $0 = \partial_t({\rm Tr}\rho_j) = {\rm Tr}(\mathcal{L}\rho_j) = \lambda_j{\rm Tr}\rho_j$.
\end{proof}

\begin{lemma}\label{lemma5}
    In the situation where $\mathcal{L}$ is not diagonalizable, for an eigenvalue $\lambda_j\neq 0$ whose geometric multiplicity is smaller than its algebraic multiplicity, all the (linearly independent) ``column vectors'' of the transition operator $\mathcal{P}$ associated with $\lambda_j$ are traceless, where $\mathcal{P}^{-1}\mathcal{L}\mathcal{P} = \mathcal{J}$ and $\mathcal{J}$ is the Jordan normal form.
\end{lemma}
\begin{proof}
    Without losing generality, we take an example of a 3-fold degenerate eigenvalue $\lambda_j$ whose geometric multiplicity is 1. Then, in this subspace we can explicitly write $\mathcal{L}_{\lambda_j}$ in matrix form:
    \beq
        \mathcal{L}_{\lambda_j} = \mathcal{P}_{\lambda_j}
        \bpm
            \lambda_j&1&0\\
            0&\lambda_j&1\\
            0&0&\lambda_j
        \epm
        \mathcal{P}_{\lambda_j}^{-1},\ \ \ \mathcal{P}_{\lambda_j} = 
        \bpm
            \rho_j&p_1&p_2
        \epm.
    \eeq
    Lemma \ref{lemma4} tells us that ${\rm Tr}\rho_j=0$. Then, we can show:
    \beq
        \partial_t({\rm Tr}p_1) = {\rm Tr}(\mathcal{L}_{\lambda_j}p_1) = {\rm Tr}\rho_j + \lambda_j {\rm Tr}p_1 = 0 \Longrightarrow {\rm Tr}p_1=0.
    \eeq
    Likewise, ${\rm Tr}p_2 = 0$ can be proved by continued induction.
\end{proof}

\begin{lemma}\label{lemma6}
    $\mathcal{L}$ must have at least one eigenstate $\rho_0$ such that $\lambda_0=0$ while ${\rm Tr}\rho_j\neq 0$. This $\rho_0$, by definition, is a steady state of the system such that $\partial_t\rho_0 = \mathcal{L}\rho_0 = 0$.
\end{lemma}
\begin{proof}
    This can be shown straightforwardly with $\mathcal{L}$ assumed to be diagonalizable (because its eigenstates form a complete basis of the operator space $\mathcal{H}\otimes\mathcal{H}^*$) and Lemma \ref{lemma4} shows that eigenstates with non-zero eigenvalues are traceless.
    
    Furthermore, without assuming diagonalizability, we can also show the existence of a zero eigenvalue and a corresponding traceful eigenoperator. As mentioned in the proof of Lemma \ref{lemma5}, in the situation where $\mathcal{L}$ is not diagonalizable, the ``column vectors'' of the transition operator $\mathcal{P}$ still form a complete basis of the operator space. Therefore, there must be at least one ``column vector'' with non-zero trace. Lemma \ref{lemma5} then already shows that all the ``column vectors'' in the $\lambda_j\neq 0$ subspaces are traceless; thus, there is at least one traceful basis state in the $\lambda_0=0$ subspace. As a consequence, we prove the existence of $\lambda_0=0$. 
    
    We are now going to prove that at least one eigenstate of $\lambda_0=0$ is traceful, case by case:
    
    1. If $\lambda_0$ is non-degenerate, the corresponding eigenstate $\rho_0$ must have non-zero trace.

    2. If in the $\lambda_0$ subspace $\mathcal{L}_{\lambda_0}$ is diagonalizable, then all the corresponding ``column vectors'' in $\mathcal{P}$ are eigenstates of $\lambda_0$. The above argument of the completeness of the basis leads to the result that at least one of them has a non-zero trace.

    3. The only remaining case is that $\mathcal{L}_{\lambda_0}$ is not diagonalizable. However, we will next show that $\mathcal{L}_{\lambda_0}$ is always diagonalizable.
\end{proof}

\begin{lemma}\label{lemma7}
    If $\lambda_0=0$ has degeneracy $n$, then there exist $n$ independent corresponding eigenvectors \cite{spectrum_proof}. Namely, $\mathcal{L}_{\lambda_0}$ is diagonalizable.
\end{lemma}
\begin{proof}
     We will prove this by contradiction. Supposing that $\mathcal{L}_{\lambda_0}$ is not diagonalizable, we can always put the non-diagonalizable part into its Jordan normal form:
    \beq
        \mathcal{L}_{\lambda_0} = \mathcal{P}_{\lambda_0}
        \bpm
            \lambda_0\ \ &1&\ &\ \\
            0\ \ &\lambda_0&1&\ \\
            \ &0&\ddots&\ddots\\
            \ &\ &\ddots&\ddots&1\\
            \ &\ &\ &0&\lambda_0
        \epm
        \mathcal{P}_{\lambda_0}^{-1}.
    \eeq
    Then we can express the time evolution operator as:
    \beq
        e^{\mathcal{L}_{\lambda_0}t} = \mathcal{P}_{\lambda_0} e^{\lambda_0 t}
        \bpm
            1\ \ &t&\frac{t^2}{2}&...&\frac{t^{(n-1)}}{(n-1)!}&\frac{t^{n}}{n!}\\
            0\ \ &1&t&...&\frac{t^{(n-2)}}{(n-2)!}&\frac{t^{n-1}}{(n-1)!}\\
            \ &0&\ddots&\ddots\\
            \ &\ &\ddots&\ddots&1&t\\
            \ &\ &\ &\ &0&1
        \epm
        \mathcal{P}_{\lambda_0}^{-1}.
    \eeq
    Since $\lambda_0=0$, the above expression will clearly cause the dynamics to diverge at large $t$, which gives a contradiction.
\end{proof}

\begin{lemma}\label{lemma8}
    The property stated in Lemma \ref{lemma7} also applies to purely imaginary eigenvalues ${\rm Re}[\lambda_j] = 0$. (Proof is similar.)
\end{lemma}

\begin{lemma}\label{lemma9}
    If $\lambda_0=0$ is non-degenerate, the corresponding $\rho_0$ is Hermitian, semi-definite, and traceful, indicating that $\rho_0$ is physical after normalizing to ${\rm Tr}\rho_0=1$.
\end{lemma}
\begin{proof}
    Lemma \ref{lemma3} and \ref{lemma6} already showed that $\rho_0$ is Hermitian and traceful. Thus, we only need to prove its semi-definiteness here, or equivalently, prove positive semi-definite after normalizing to ${\rm Tr}\rho_0=1$. 

    We prove this by construction. We take a physical density operator as the initial state and let it evolve. Since the ``column vectors'' of $\mathcal{P}$ form a complete basis, the initial state can be expressed as a linear combination of them. At a sufficiently long time $t$, all terms in the ${\rm Re}[\lambda_j]<0$ subspaces decay due to the $e^{\lambda_jt}$ factor in the evolution operator $e^{\mathcal{L}_{\lambda_j}t}$, leaving only the components in the ${\rm Re}[\lambda_j]=0$ subspaces.

    Then, Lemma \ref{lemma7} and \ref{lemma8} show that these subspaces are fully expanded by the corresponding eigenvectors of $\mathcal{L}$. Therefore, at large time $t$, the evolved physical state, which is Hermitian, can be expressed as
    \beq
        \rho(t) = a_0\rho_0 + \sum_{j:{\rm Re}[\lambda_j]=0,{\rm Im}[\lambda_j]\equiv\omega_j\neq 0}(e^{i\omega_jt}a_j\rho_j+h.c.),\ \ \ a_0\in\mathbb{R}.
    \eeq
    The CP (completely positive) nature of the dynamical evolution ensures that the above remains positive semi-definite, which means that 
    \beq
        \forall \ket{\psi}\in\mathscr{H},\ \bra{\psi}\rho(t)\ket{\psi}\geq 0.
    \eeq
    Then, denoting $\bra{\psi}\rho_j\ket{\psi}=\braket{\rho_j}$, we have
    \beq
        a_0\braket{\rho_0} + \sum_{j:{\rm Re}[\lambda_j]=0,{\rm Im}[\lambda_j]\equiv\omega_j\neq 0}(e^{i\omega_jt}a_j\braket{\rho_j}+c.c.)\geq0.
    \eeq
    The terms in the summation must take the form $\tilde{a}_j\cos(\omega_jt+\phi_j),\ \ \tilde{a}_j\in\mathbb{R}$, so that we have 
    \beq
        a_0\braket{\rho_0} + \sum_{j:{\rm Re}[\lambda_j]=0,{\rm Im}[\lambda_j]\equiv\omega_j\neq 0}\tilde{a}_j\cos(\omega_jt+\phi_j)\geq0.
    \eeq
    Now we see that the summation is over a finite set of sinusoidal functions. This function can be shown to be “almost periodic” and has infinitely many zeros. Here, we only use the fact that its long-time average is zero, which indicates that it is sometimes positive and sometimes negative, unless all the $a_j$’s are zero (in that case the summation is zero). As a result, the first term $a_0\braket{\rho_0}$ should be non-negative to keep the entire equation non-negative.

    Therefore, $\bra{\psi}\rho_0\ket{\psi}$ is either non-positive or non-negative for all $\ket{\psi}\in\mathscr{H}$, and thus $\rho_0$ is semi-definite. After normalizing to ${\rm Tr}\rho_0=1$, it is clearly positive semi-definite.
\end{proof}

\begin{lemma}\label{lemma10}
    Regardless of whether $\lambda_0=0$ is degenerate or non-degenerate, there exists at least one physical steady state which is an eigenoperator of $\lambda_0$.
\end{lemma}
\begin{proof}
    This is similar to the previous proof. Starting from a physical initial state and evolving for a long enough time such that all transients have decayed, the late-time state is
    \beq
    \begin{aligned}
        \rho(t) &=\sum_{n} (a_0^{(n)}\rho_0^{(n)} + a_0^{*(n)}\rho_0^{(n)\dag})\\
        &+ \sum_{j:{\rm Re}[\lambda_j]=0,{\rm Im}[\lambda_j]\neq 0}(e^{i\omega_jt}a_j\rho_j+h.c.),
    \end{aligned}
    \eeq
    where $\rho_0^{(n)}$ are the eigenoperators with eigenvalue $\lambda_0$. Now, we see that the first term is Hermitian and must have a unit trace, since the second term has vanishing trace. A similar argument as one used in the proof of Lemma~\ref{lemma9} shows that the expectation value of the first term under any state $\ket{\psi}\in\mathscr{H}$ in the Hilbert space is positive.
    Now we have shown that the first term is physical. Noticing that the first term is just a steady state, we conclude that there is at least one physical steady state of the system.
\end{proof}

In summary, the non-decaying part of the time-evolved state $\rho(t)$ from any arbitrary initial state $\rho(t_0)$ can be expressed as a linear combination of eigenstates $\rho_j$ of $\mathcal{L}$ whose eigenvalues have zero real parts ${\rm Re}[\lambda_j] = 0$. 


\section{Floquet theory for closed quantum systems}
\label{app:floquet}

Here, we review the Floquet theory for closed quantum systems evolving under a time-periodic Hamiltonian. Consider the time-dependent Schr\"{o}dinger equation:
\beq
    i\frac{d}{dt}\ket{\psi} = H(t)\ket{\psi},
\eeq
where the Hamiltonian $H(t) = H(t+T)$ is periodic in time. It is well known that the whole dynamics is captured by the unitary time-evolution operator
\beq
    U(t,t_0) = \mathcal{T}e^{-i\int_{t_0}^t H(t)dt},
\eeq
and the discrete time-translational invariance lets us break the entire time interval into several periods plus a residual time: $t-t_0 = mT+\tau$, $m\in\mathbb{N}_0$ and $\tau\in[0,T)$. As a result:
\beq
    U(t,t_0) = U(t_0+\tau,t_0)(U(t_0+T,t_0))^m.
\eeq
The evolution operator over one period is defined as the \emph{Floquet evolution operator}:
\beq
    U_F\equiv U(t_0+T,t_0) = \mathcal{T}e^{-i\int_{t_0}^{t_0+T} H(t)dt},
\eeq
which is usually written in the form $U_F \equiv e^{-iH_FT}$, where $H_F$ is a Hermitian operator referred to as the  ``Floquet Hamiltonian". $H_F$ is not a Hamiltonian in any physical sense since it is generically non-local. In addition, it is not unambiguously defined due to the indeterminacy modulo $2\pi $ on its eigenvalues.

Since $U_F$ is unitary, its diagonalization provides a complete and orthonormal basis $\{\ket{u_j}\}$ of the Hilbert space $\mathscr{H}$:
\beq
    U_F\ket{u_j} = e^{-i\epsilon_jT}\ket{u_j}
\eeq
where the ``quasi-energies" $\epsilon_j$ are determined only modulo $\Omega=2\pi/T$. Note that these $\ket{u_j}$'s are \textit{time-independent}. Clearly, these eigenstates are also the eigenstates of $H_F$:
\beq
    H_F\ket{u_j} = \epsilon_j\ket{u_j}.
\eeq

Calculating the time-evolution of these eigenstates $\ket{u_j}$, we get
\beq
\begin{aligned}
    U(t,t_0)\ket{u_j} &= U(t_0+\tau,t_0)(U_F)^m\ket{u_j}\\
    &= e^{-i\epsilon_jmT}U(t_0+\tau,t_0)\ket{u_j},
\end{aligned}
\eeq
which can be seen as an envelope $e^{i\epsilon_jmT}$ modulating a periodic function $U(t_0+\tau,t_0)\ket{u_j}$. This is not satisfactory since both of them are not continuous at the end of each period $\tau=T$, so we fix it by introducing an extra factor of $e^{-i\epsilon_j\tau}$ into the envelope:
\beq
\begin{aligned}
    U(t,t_0)\ket{u_j} &= e^{-i\epsilon_j(mT+\tau)}U(t_0+\tau,t_0)e^{i\epsilon_j\tau}\ket{u_j}\\
    &= e^{-i\epsilon_j(t-t_0)}\ket{u_j(t)}.
\end{aligned}
\eeq
The periodic part is defined as Floquet (eigen)modes: $\ket{u_j(t)} = U(t_0+\tau,t_0)e^{i\epsilon_j\tau}\ket{u_j}$, or in a neater but equivalent way: $\ket{u_j(t)} = U(t,t_0)e^{i\epsilon_j(t-t_0)}\ket{u_j}$, which are generalizations of energy eigenstates in time-independent problems. We can construct general solutions of the time-dependent Schr\"{o}dinger equation by linearly combining the Floquet modes:
\beq
    \ket{\psi(t)} = \sum_j a_j e^{-i\epsilon_j(t-t_0)}\ket{u_j(t)}.
\eeq


\section{Time-evolution of Floquet-Lindblad eigenmodes}
\label{app:evol_floquet_modes}

In this Appendix, we derive the time evolution of the Floquet-Lindblad eigenoperators $\rho_j(t)$. Taking $\rho_j$ as the initial state and applying the time evolution operator to it, we get
\beq
    \mU(t,t_0)\rho_j = \mU(t_0+\tau,t_0)(U_F)^m\rho_j = e^{\lambda_jmT}\mU(t_0+\tau,t_0)\rho_j.
\eeq
It is clear that the evolution contains a periodic part $\mU(t_0+\tau,t_0)\rho_j$ with an envelope $e^{\lambda_jmT}$. However, we are still one step away from the concept of Floquet mode, because both the periodic part and the envelope are not continuous at the end of each period $\tau=T$. To resolve this, we rewrite it in the form:
\beq
\begin{aligned}
    \mU(t,t_0)\rho_j &= e^{\lambda_jmT}\mU(t_0+\tau,t_0)e^{\lambda_j\tau}e^{-\lambda_j\tau}\rho_j \\
    &= e^{\lambda_j(mT+\tau)}\mU(t_0+\tau,t_0)e^{-\lambda_j\tau}\rho_j.
\end{aligned}
\eeq
Re-written thusly, the envelope $e^{\lambda_j(mT+\tau)} = e^{\lambda_j(t-t_0)}$ is manifestly continuous, while the remainder
\beq
    \varrho_j(t_0+\tau)\equiv \mU(t_0+\tau,t_0)e^{-\lambda_j\tau}\rho_j,\ \ \varrho_j(t+T) = \varrho_j(t),
\eeq
is also now continuous and remains periodic, similarly to the Floquet eigenmodes in non-dissipative Floquet systems. We can write $\varrho_j(t)$ in another form:
\beq
\varrho_j(t) = e^{-\lambda_j(t-t_0)}\mU(t,t_0)\rho_j.
\eeq
We refer to $\varrho_j(t)$ as the Floquet-Lindblad eigenmodes. Please note that $\varrho_j(t)$ is not $\rho_j$ evolved to time $t$: $\varrho_j(t)\neq \mU(t,t_0)\rho_j$.


\section{Shirley-Floquet Formalism and the Extended Hilbert Space}
\label{app:SF}

Generically, it is not practical to directly obtain an exact expression for $U_F$. Therefore, we need a more accessible way for obtaining the Floquet modes (as functions of time). In this Appendix, we discuss one such approach for obtaining the stroboscopic evolution operator in closed quantum systems.

From Appendix \ref{app:floquet} we already know that the time-independent $\ket{u_j}$'s (equivalently, the Floquet modes at initial time $t_0$, or more generally speaking, at any fixed time $t$) form a complete and orthonormal basis of the finite-dimensional Hilbert space $\mathscr{H}$ of the quantum system. With the help of Fourier's theorem, we know that the time-dependent Floquet modes each have a period $T = 2\pi/\Omega$ such that each of them can be expressed as (we use $e^{-il\Omega t}$ instead of $e^{-il\Omega(t-t_0)}$ here so that the effects of the initial time $t_0$ are absorbed in the coefficients $a_{jk}^{(l)}$)
\beq
\begin{aligned}
    \ket{u_j(t)} &= \sum_{k} a_{jk}(t)\ket{\psi_k}
    = \sum_{k} \left(\sum_{l=-\infty}^{\infty}a_{jk}^{(l)}e^{-il\Omega t}\right)\ket{\psi_k},
\end{aligned}
\eeq
where $\{\ket{\psi_k}\}$ forms a complete and orthonormal basis of $\mathscr{H}$ that may or may not be the same as $\{\ket{u_j}\}$. Realizing that the harmonics $\varphi^{(l)}(t) = e^{-il\Omega t}$ of a fundamental frequency $\Omega$ form a complete and orthogonal basis of another Hilbert space $\mathscr{H}_T$ of square-integrable $T$-periodic functions, with inner product defined as 
\beq
    \braket{\varphi^{(l')},\varphi^{(l)}} \equiv \frac{1}{T}\int_{t_0}^{t_0+T}dt\varphi^{(l')*}(t)\varphi^{(l)}(t) = \delta_{l,l'},
\eeq
we readily see that the time-dependent Floquet modes live in an \emph{extended Hilbert space} $\mathscr{H}_T\otimes\mathscr{H}$ (sometimes also referred to as the Sambe space) whose basis can be constructed by the direct product of $\varphi^{(l)}(t)$'s and $\ket{\psi_k}$'s: $\ket{\psi_k^{(l)}(t)} \equiv \varphi^{(l)}(t)\ket{\psi_k}$. In this basis, we can express the Floquet modes as
\beq
    \ket{u_j(t)} = \sum_{kl}a_{jk}^{(l)}\varphi^{(l)}(t)\ket{\psi_k} = \sum_{kl}a_{jk}^{(l)}\ket{\psi_k^{(l)}(t)}.
\eeq
Moreover, the periodically time-dependent Hamiltonian can also be expressed in terms of this basis:
\beq
\begin{aligned}
    H(t) &= \sum_{k,k',l,l'}h_{kk'}^{(l,l')}\ket{\psi_{k}^{(l)}(t)}\bra{\psi_{k'}^{(l')}(t)}\\
    &= \sum_{kk',(l-l')}h_{kk'}^{(l-l')}e^{-i(l-l')\Omega t}\ket{\psi_k}\bra{\psi_{k'}}\\
    H(t)&= \sum_{kk',l}h_{kk'}^{(l)}e^{-il\Omega t}\ket{\psi_k}\bra{\psi_{k'}},
\end{aligned}
\eeq
where $h_{kk'}^{(l-l')} = \frac{1}{T}\int_{t_0}^{t_0+T}dte^{i(l-l')\Omega t}\bra{\psi_k}H(t)\ket{\psi_k'}$ are the matrix elements of $H(t)$ in the extended Hilbert space. For convenience in the upcoming discussions, we denote the bold letter $\bm{h}^{(l)}$ as the matrix whose $(k,k')$ element is $h_{kk'}^{(l)}$.

Having introduced the notion of the extended Hilbert space, we now discuss the Shirley-Floquet formalism. First of all, since $U(t,t_0)\ket{u_j} = e^{-i\epsilon_j(t-t_0)}\ket{u_j(t)}$ satisfies the Schr\"{o}dinger equation, it is clear that the Floquet modes $\ket{u_j(t)}$ satisfy:
\beq
\begin{aligned}
    i\frac{d}{dt}(e^{-i\epsilon_j t}\ket{u_j(t)}) &= \epsilon_j e^{-i\epsilon_j t}\ket{u_j(t)} + e^{-i\epsilon_j t}i\frac{d}{dt}\ket{u_j(t)}\\
    &= H(t)e^{-i\epsilon_j t}\ket{u_j(t)}
\end{aligned}
\eeq
and thus 
\begin{equation}
    \left(H(t) - i\frac{d}{dt}\right)\ket{u_j(t)} = \epsilon_j\ket{u_j(t)}. 
\end{equation}
That is, the periodically time-dependent Floquet mode $\ket{u_j(t)}$ is the eigenstate of the operator $K(t)\equiv H(t)-i\frac{d}{dt}$ acting on the extended Hilbert space, with eigenvalue $\epsilon_j$. Note that the action of $K(t)$ on the extended Hilbert space is crucial, else any solution of the time-dependent Schr\"{o}dinger equation would be an eigenstate of $K(t)$ with zero eigenvalue. Here, we emphasize again that the quasi-energy $\epsilon_j$ is not fully determined: if we consider a closely related Floquet mode $e^{il\Omega(t-t_0)}\ket{u_j(t)}$, it can be verified that $K(t)[e^{il\Omega(t-t_0)}\ket{u_j(t)}] = (\epsilon_j+l\Omega)e^{il\Omega(t-t_0)}\ket{u_j(t)}$. This result indicates that multiplying the Floquet mode $\ket{u_j(t)}$ by $e^{il\Omega(t-t_0)}$ leads to a different eigenvalue $\epsilon_j + l\Omega$. This will lead to some redundancy in the Floquet modes obtained through the Shirley-Floquet formalism.

The equation for solving Floquet modes in Shirley-Floquet formalism is the above eigenvalue equation in the extended Hilbert space, which is usually written in the product basis:
\beq
\begin{aligned}
    H(t)\ket{u_j(t)} &= \sum_{kk',l'}h_{kk'}^{(l')}e^{-il'\Omega t}\ket{\psi_k}\bra{\psi_{k'}}\\
    &\times\sum_{k'',l''}a_{jk''}^{(l'')}e^{-il''\Omega t}\ket{\psi_{k''}}\\
    &= \sum_{kk'k'',l'l''}h_{kk'}^{(l')}e^{-i(l'+l'')\Omega t}a_{jk''}^{(l'')}\ket{\psi_k}\braket{\psi_{k'}|\psi_{k''}}\\
    &= \sum_{kk',l'l''}h_{kk'}^{(l')}e^{-i(l'+l'')\Omega t}a_{jk'}^{(l'')}\ket{\psi_k}\\
    &= \sum_{kk',ll''}h_{kk'}^{(l-l'')}a_{jk'}^{(l'')}e^{-il\Omega t}\ket{\psi_k}, \quad (l'+l''\equiv l)\\
    -i\frac{d}{dt}\ket{u_j(t)}
    &= \sum_{k,l}-l\Omega a_{jk}^{(l)}e^{-il\Omega t}\ket{\psi_{k}}.
\end{aligned}
\eeq
Matching the coefficients in front of each basis state $\ket{\psi_k^{(l)}(t)}=e^{-il\Omega t}\ket{\psi_k}$, we get the equation for each of the coefficients:
\beq\label{eq:SF1}
    \sum_{k',l'}h_{kk'}^{(l-l')}a_{jk'}^{(l')} - l\Omega a_{jk}^{(l)} = \epsilon_j a_{jk}^{(l)},
\eeq
or more concisely:
\beq\label{eq:SF2}
    \sum_{k',l'}\left(h_{kk'}^{(l-l')}-l\Omega\delta_{kk'}\delta_{ll'}\right)a_{jk'}^{(l')} = \epsilon_j a_{jk}^{(l)}.
\eeq

By solving the above equations, one can obtain the coefficients $a_{jk}^{(l)}$ in the product basis of the extended Hilbert space, which physically correspond to the Fourier coefficients in the harmonic mode $l\Omega$ and the quantum Hilbert space basis state $\ket{\psi_j}$. We refer to this as the Schr\"{o}dinger equation in \emph{Shirley-Floquet (SF) form}. Rewritten in matrix form, the operator $K(t)$ acts on the extended Hilbert space $\mathscr{H}_T\otimes\mathscr{H}$ and its matrix elements in the direct product basis $\{\ket{\psi_k^{(l)}} = \varphi^{(l)}(t)\ket{\psi_k}\}$ are given by
\beq\label{eq:SFH}
    \bra{\psi_k^{(l)}}K(t)\ket{\psi_{k'}^{(l')}} = h_{kk'}^{(l-l')} - l\Omega\delta_{kk'}\delta_{ll'}\equiv (\bm{h}_{SF})_{kk'}^{(l-l')}.
\eeq
This matrix $\bm{h}_{SF}$ is typically referred to as the \textit{Shirley-Floquet Hamiltonian}, which is infinite-dimensional. This is because we now associate each basis state $\ket{\psi_k}$ with infinitely many Fourier coefficients with indices $l = 0, \pm 1, \pm 2,...$. As a result, the eigensolutions contain infinitely many eigenvalues and corresponding eigenstates that are linearly independent in the extended Hilbert space. This reveals the redundancy we mentioned above: starting with one Floquet mode $\ket{u_j(t)}$ with quasi-energy $\epsilon_j$, we can multiply it by $e^{il\Omega(t-t_0)}$ with any arbitrary integer $l$ to generate another Floquet mode with quasi-energy $\epsilon_j + l\Omega$. This procedure leads to infinitely many eigenvalues and eigenstates in the SF form (the quasi-energy $\epsilon_j$ is restricted to lie between 0 and $\Omega$.) 

We see that by implementing the SF formalism, say for instance in a system that originally has $n$ energy levels and is driven harmonically, we obtain (infinitely) many Floquet energy levels. These are the original $n$ levels shifted up or down by multiples of $\Omega$, which correspond to the $l\Omega$ terms on the diagonal of the SF Hamiltonian; the driving can thus be thought of as effectively allowing ``hopping" between different Floquet levels. This perspective, which is often used in the condensed matter literature, coincides with the tight-binding form of (\ref{eq:SF2}) and (\ref{eq:SFH}). Explicitly, we can place the different Fourier components of $\bm{h}_{SF}$ in a large matrix:
\beq\label{eq:SFHmatrix}
\begin{aligned}
    \bm{h}_{SF} &\equiv 
    \bpm
        \ddots & \vdots & \vdots & \vdots & \ddots\\
        \hdots & \bm{h}^{(0)} + \Omega\bm{I} & \bm{h}^{(-1)} & \bm{h}^{(-2)} & \hdots\\
        \hdots & \bm{h}^{(1)} & \bm{h}^{(0)} & \bm{h}^{(-1)} & \hdots\\
        \hdots & \bm{h}^{(2)} & \bm{h}^{(1)} & \bm{h}^{(0)} - \Omega\bm{I} & \hdots\\
        \ddots & \vdots & \vdots & \vdots & \ddots
    \epm
    \\
    &= \bm{h} + 
    \bpm
        \ddots\\
        \ & \Omega\bm{I} & \ &\ & \ \\
        \ & \ & \bm{0} & \ & \ \\
        \ & \ & \ & -\Omega\bm{I} & \ \\
        \ & \ & \ & \ & \ddots
    \epm
\end{aligned}
\eeq
and the vector $\bm{a}_j$ can be written in a corresponding way:
\beq
    \bm{a}_j \equiv 
    \bpm
        \vdots\\
        \bm{a}_j^{(-1)}\\
        \bm{a}_j^{(0)}\\
        \bm{a}_j^{(1)}\\
        \vdots
    \epm
    =
    \bpm
        \vdots\\
        
        a_{j1}^{(-1)}\\
        a_{j2}^{(-1)}\\
        \vdots\\
        a_{jn}^{(-1)}\\
        a_{j1}^{(0)}\\
        a_{j2}^{(0)}\\
        \vdots\\
        a_{jn}^{(0)}\\
        a_{j1}^{(1)}\\
        a_{j2}^{(2)}\\
        \vdots\\
        a_{jn}^{(1)}\\
        \vdots
    \epm.
\eeq
Then, Eq.~(\ref{eq:SF2}) can be rewritten as
\beq
    \bm{h}_{SF}\bm{a}_j = \epsilon_j \bm{a}_j \ .
\eeq
We will find it useful to represent $\bm{h}_{SF}$ in the explicit form given in Eq.~(\ref{eq:SFHmatrix}) when generalizing our discussion to the Floquet-Lindblad equation, where we will have to work with density matrices instead.

To implement the SF formalism in practice, we need to place cutoffs on these infinite dimensional matrices, preserving only certain elements in $\bm{h}_{SF}$ and $\bm{a}$. Typically, when performing numerics, one usually preserves all the components in $\bm{a}$ within a given frequency range: $l = 0,\pm 1,\pm 2,...,\pm L$, such that $\bm{h}_{SF}$ is also approximated by a $(2L+1)N\times(2L+1)N$ square matrix, where $N$ is the dimension of $\mathscr{H}$. This $(2L+1)N\times(2L+1)N$ square matrix sits in the middle of the infinitely large $\bm{h}_{SF}$. 

Besides placing the cutoff, in dealing with two-level systems, the rotation wave approximation (RWA) is also often used as a first-order approximation. Here, the RWA means that we keep only two coefficients in $\bm{a}$: $a_1^{(-1)}$ and $a_2^{(0)}$, or equivalently speaking, we are only interested in the sub-extended Hilbert space spanned by $\{\ket{\psi_1^{(-1)}},\ket{\psi_2^{(0)}}\}$. As a result, the matrix $\bm{h}_{SF}$ is approximated by a 2 by 2 matrix:
\beq
    \bm{h}^{\rm RWA}_{SF} \equiv\bpm
        h_{11}^{(0)} + \Omega & h_{12}^{(-1)}\\
        h_{21}^{(1)} & h_{22}^{(0)}
    \epm.
\eeq


\section{Shirley-Floquet formalism for Floquet-Lindblad equations}
\label{app:SFL}

Following our discussion of the SF formalism in closed quantum systems, we now extend it to open quantum systems, specifically those described by the Floquet-Lindblad master equation. For simplicity, we will assume here that the jump operators and their associated decay rates are time-independent, with the periodic time dependence stemming purely from the coherent Hamiltonian. Imitating the procedure from the preceding Appendix, we first bring the Floquet-Lindblad equation into the form:
\beq
    \left(\mathcal{L}(t) - \frac{d}{dt}\right)\rho(t) \equiv \mathcal{K}(t)\rho(t) = 0,
\eeq
where $\rho(t)$ is some solution. However, if we consider a Floquet-Lindblad eigenmode $\varrho_j(t)$, it is \textit{not} a solution of the above equation: the proper solution is instead given by $\mU(t,t_0)\rho_j = e^{\lambda_j(t-t_0)}\varrho_j(t)$, which leads to
\beq
\begin{aligned}
    &\left(\mathcal{L}(t)-\frac{d}{dt}\right)e^{\lambda_j(t-t_0)}\varrho_j(t) = 0\\
\end{aligned}
\eeq
and hence $\mathcal{K}(t)$ has the same spectrum ${\lambda_j}$,
\begin{equation}
    \mathcal{K}(t)\varrho_j(t) = \lambda_j\varrho_j(t).
\end{equation}
Therefore, we see that similarly to closed quantum systems, the periodically time-dependent Floquet-Lindblad eigenmode $\varrho_j(t)$ is an eigenoperator of the superoperator $\mathcal{K}(t)$ acting on the \textit{extended Liouville space}. A similar SF formalism can then be established for the open systems.

Again, since the Floquet-Lindblad eigenmodes $\varrho_j(t)$'s are time-periodic, we can expand them in the product basis, as we did for the time-dependent unitary case:
\beq
\begin{aligned}
    \varrho_j(t) &= \sum_{k,k',l,l'}\varrho_{j,kk'}^{(l,l')}\ket{\psi_{k}^{(l)}(t)}\bra{\psi_{k'}^{(l')}(t)}\\
    &= \sum_{kk',(l-l')}\varrho_{j,kk'}^{(l-l')}e^{-i(l-l')\Omega t}\ket{\psi_k}\bra{\psi_{k'}}\\
    \varrho_j(t)&= \sum_{kk',l}\varrho_{j,kk'}^{(l)}e^{-il\Omega t}\ket{\psi_k}\bra{\psi_{k'}}.
\end{aligned}
\eeq
The jump operators are time-independent, which can be easily expressed in the product basis with only zero frequency terms:
\beq
    L_\mu = \sum_{kk'}L_{\mu,kk'}\ket{\psi_k}\bra{\psi_{k'}} \equiv \sum_{kk'}L_{\mu,kk'}^{(0)}\ket{\psi_k}\bra{\psi_{k'}},
\eeq
where the superscript $(0)$ denotes zero frequency. As in the previous Appendix, we will use bold letters ($\bm{h}^{(l)}$ and $\bm{\varrho}_j^{(l)}$) to represent matrices. 

Using the above, we can now express the equation for calculating $\varrho_j(t)$ in the Shirley-Floquet formalism. First of all, consider the Hamiltonian term and the time-derivative $d/dt$: 
\beq
\begin{aligned}
    -i[H(t),\varrho_j(t)] &= -i\sum_{ll'}e^{-i(l+l')\Omega t}\sum_{kk'}\left[\bm{h}^{(l)},\bm{\varrho}_{j}^{(l')}\right]_{kk'}\ket{\psi_k}\bra{\psi_{k'}}\\
    &= -i\sum_{ll'}e^{-il\Omega t}\sum_{kk'}\left[\bm{h}^{(l-l')},\bm{\varrho}_{j}^{(l')}\right]_{kk'}\ket{\psi_k}\bra{\psi_{k'}},\\
    -\frac{d}{dt}\varrho_j(t) &= i\sum_{l'}l'\Omega e^{-il'\Omega t}\sum_{kk'}\varrho_{j,kk'}^{(l')}\ket{\psi_k}\bra{\psi_{k'}}\\
    &= i\sum_{l'l''}l'\Omega e^{-i(l'+l'')\Omega t}\delta_{l''0}\sum_{kk'}\varrho_{j,kk'}^{(l')}\ket{\psi_k}\bra{\psi_{k'}}\\
    (l\equiv l'+l'')&= i\sum_{ll'}l'\Omega e^{-il\Omega t}\delta_{ll'}\sum_{kk'}\varrho_{j,kk'}^{(l')}\ket{\psi_k}\bra{\psi_{k'}}\\
    &=  i\sum_{ll'}l\Omega e^{-il\Omega t}\delta_{ll'}\sum_{kk'}\varrho_{j,kk'}^{(l')}\ket{\psi_k}\bra{\psi_{k'}}\\
    &= i\sum_{ll'}e^{-il\Omega t}\sum_{kk'}l\Omega\delta_{ll'}\varrho_{j,kk'}^{(l')}\ket{\psi_k}\bra{\psi_{k'}}.
\end{aligned}
\eeq
Combining these two parts we obtain
\begin{widetext}
\beq
    -i[H(t),\varrho_j(t)] - \frac{d}{dt}\varrho_j(t)
    = -i\sum_{ll'}e^{-il\Omega t}\sum_{kk'}\left(\left[\bm{h}^{(l-l')},\bm{\varrho}_{j}^{(l')}\right]_{kk'} - l\Omega\delta_{ll'}\varrho_{j,kk'}^{(l')}\right)\ket{\psi_k}\bra{\psi_{k'}}.
\eeq
\end{widetext}
Next, we consider the dissipator, which only depends on time through $\varrho_j(t)$. We realize that in this part, the matrix elements of the operators are placed in a same way as the operators themselves:
\begin{widetext}
\beq
L_\mu \varrho_j(t) L_\mu^\dagger - \frac{1}{2}\{L_\mu^\dagger L_\mu,\varrho_j(t)\} = \sum_{l}e^{-il\Omega t}\sum_{kk'} \left(\bm{L}_\mu^{(0)}\bm{\varrho}_j^{(l)}\bm{L}_\mu^{(0)\dag} - \frac{1}{2}\left\{\bm{L}_\mu^{(0)\dag} \bm{L}_\mu^{(0)},\bm{\varrho}_j^{(l)}\right\}\right)_{kk'}\ket{\psi_k}\bra{\psi_{k'}},
\eeq
\end{widetext}
where $\bm{L}_\mu^{(0)}$ denotes the matrix whose matrix elements are $(\bm{L}_{\mu}^{(0)})_{kk'} = \bra{\psi_k}L_\mu\ket{\psi_{k'}}$. As a result, the equation for calculating $\varrho_j(t)$ in the SF formalism can be expressed through the matrix elements:
\begin{widetext}
\beq\label{eq:SFLmatrix}
    -i\sum_{l'}\left(\left[\bm{h}^{(l-l')},\bm{\varrho}_{j}^{(l')}\right] - l\Omega\delta_{ll'}\bm{\varrho}_{j}^{(l')}\right) + \sum_\mu\Gamma_\mu  \left(\bm{L}_\mu^{(0)}\bm{\varrho}_j^{(l)}\bm{L}_\mu^{(0)\dag} - \frac{1}{2}\left\{\bm{L}_\mu^{(0)\dag} \bm{L}_\mu^{(0)},\bm{\varrho}_j^{(l)}\right\}\right) = \lambda_j\bm{\varrho}_j^{(l)}.
\eeq
\end{widetext}

While the above equation is already feasible for numerical calculations, we can further simplify it to hide the $- l\Omega\delta_{ll'}\bm{\varrho}_{j}^{(l')}$ term using an analog of $\bm{h}_{SF}$ (introduced in the previous Appendix). It can be shown that the commutator between the matrices $\bm{h}_{SF}$ and $\bm{\varrho}$ correctly captures the information encoded in $\left[\bm{h}^{(l-l')},\bm{\varrho}_{j}^{(l')}\right] - l\Omega\delta_{ll'}\bm{\varrho}_{j}^{(l')}$:
\beq
\begin{aligned}
    &[\bm{h}_{SF},\bm{\varrho}_j]\\
    =& [\bm{h},\bm{\varrho}_j] + 
    \bpm
        \ddots & \vdots & \vdots & \vdots & \ddots\\
        \hdots & \bm{0} & \Omega\bm{\varrho}^{(-1)}_j & 2\Omega\bm{\varrho}^{(-2)}_j & \hdots\\
        \hdots & -\Omega\bm{\varrho}^{(1)}_j & \bm{0} & \Omega\bm{\varrho}^{(-1)}_j & \hdots\\
        \hdots & -2\Omega\bm{\varrho}^{(2)}_j & -\Omega\bm{\varrho}^{(1)}_j & \bm{0} & \hdots\\
        \ddots & \vdots & \vdots & \vdots & \ddots
    \epm.
\end{aligned}
\eeq
As a result, the Shirley-Floquet equation to solve the Floquet-Lindblad eigenmodes can be written in the following compact form
\beq\label{eq:SFLcompact}
    -i[\bm{h}_{SF},\bm{\varrho}_j] + \sum_\mu \Gamma_\mu \left(\bm{L}_\mu\bm{\varrho}_j\bm{L}_\mu^{\dag} - \frac{1}{2}\left\{\bm{L}_\mu^{\dag} \bm{L}_\mu^{},\bm{\varrho}_j\right\}\right) = \lambda_j\bm{\varrho}_j,
\eeq
which looks the same as the operator equation, and
\beq
\begin{aligned}
    \bm{L}_\mu &= \bpm
        \ddots & \vdots & \vdots & \vdots & \ddots\\
        \hdots & \bm{L}_\mu^{(0)} & \bm{0} & \bm{0} & \hdots\\
        \hdots & \bm{0} & \bm{L}_\mu^{(0)} & \bm{0} & \hdots\\
        \hdots & \bm{0} & \bm{0} & \bm{L}_\mu^{(0)} & \hdots\\
        \ddots & \vdots & \vdots & \vdots & \ddots
    \epm,\\
    \bm{\varrho}_j &= \bpm
        \ddots & \vdots & \vdots & \vdots & \ddots\\
        \hdots & \bm{\varrho}^{(0)}_j & \bm{\varrho}^{(-1)}_j & \bm{\varrho}^{(-2)}_j & \hdots\\
        \hdots & \bm{\varrho}^{(1)}_j & \bm{\varrho}^{(0)}_j & \bm{\varrho}^{(-1)}_j & \hdots\\
        \hdots & \bm{\varrho}^{(2)}_j & \bm{\varrho}^{(1)}_j & \bm{\varrho}^{(0)}_j & \hdots\\
        \ddots & \vdots & \vdots & \vdots & \ddots
    \epm.
\end{aligned}
\eeq

For the specific two-band model we consider in the main text, which is effectively a two-level system in each momentum sector, the band indices are given by $k=1,2$ for the valance and conduction bands respectively. The Floquet index is $l = 0,\pm 1,\pm 2,...$, and the jump operators are $\bm{L}_1^{(0)} = \sigma^+$ and $\bm{L}_2^{(0)} = \sigma^-$. In this context, making the RWA means that we are interested in the matrix elements only for the 2D sub-extended Hilbert space spanned by $\{\ket{\psi_1^{(-1)}},\ket{\psi_2^{(0)}}\}$. For an arbitrary operator $O(t)$ that has periodic time dependence, we are interested in the following four elements (the subscript ``red'' denotes ``reduced'')
\beq
    O(t)\longrightarrow \bm{O}_{\rm red}(t) \equiv
    \bpm
        O_{11}^{(0)} & O_{12}^{(-1)}e^{i\Omega t}\\
        O_{21}^{(1)}e^{-i\Omega t} & O_{22}^{(0)}
    \epm,
\eeq
where $O_{kk'}^{(l-l')} \equiv \bra{\psi_k^{(l)}}O(t)\ket{\psi_{k'}^{(l')}} = \frac{1}{T} \int_{t_0}^{t_0+T} dt e^{i(l-l')\Omega t}\bra{\psi_k}O(t)\ket{\psi_{k'}}$, such that $O_{kk'}(t) = \sum_{l}O_{kk'}^{(l)}e^{-il\Omega t}$. In other words, we approximate most of the matrix elements as zero, except for the following four elements:
\beq
    \bm{\varrho}_j^{\rm RWA}\equiv\bpm
        \varrho_{j,11}^{(0)} & \varrho_{j,12}^{(-1)}\\
        \varrho_{j,21}^{(1)} & \varrho_{j,22}^{(0)} 
    \epm,
\eeq
which can be solved using the matrix equation Eq.~(\ref{eq:SFLmatrix}). This means that we can also keep only the corresponding four elements in $\bm{h}$:
\beq
    \bm{h}^{\rm RWA} \equiv\bpm
        h_{11}^{(0)} & h_{12}^{(-1)}\\
        h_{21}^{(1)} & h_{22}^{(0)} 
    \epm.
\eeq
As a result, under this approximation Eq.~(\ref{eq:SFLmatrix}) reduces to
\begin{widetext}

\beq
    -i\left(\left[\bm{h}^{\rm RWA},\bm{\varrho}_j^{\rm RWA}\right] -\Omega
    \bpm
        0 & -\varrho_{j,12}^{(-1)}\\
        \varrho_{j,21}^{(1)} & 0
    \epm\right)
     + \sum_{\mu = 1}^{2}\Gamma_\mu  \left(\bm{L}_\mu\bm{\varrho}_j^{\rm RWA}\bm{L}_\mu^\dag - \frac{1}{2}\left\{\bm{L}_\mu^\dagger \bm{L}_\mu,\bm{\varrho}_j^{\rm RWA}\right\}\right) = \lambda_j\bm{\varrho}_j^{\rm RWA},
\eeq  
\end{widetext}
which can be rewritten into the compact form Eq. (\ref{eq:SFLcompact}) by using $\bm{h}^{\rm RWA}_{SF}$. 


\section{Evaluation of nonlinear optical responses}
\label{app:nonlinearopt}

In this Appendix, we evaluate the non-linear optical effect in a two-band model as an example to illustrate the procedure of evaluating physical quantities within the Floquet-Lindblad formalism. The traditional setup is the following: we shine a spatially uniform beam of laser on a piece of solid with intrinsic (original) Hamiltonian $H^0$, rendering the system Hamiltonian time-dependent:
\beq\label{eq:nonlinear_hamiltonian}
    H(t) = H^0 + A(t)V^0,
\eeq
where $A(t)$ is the vector potential and we have kept only the linear term in $A(t)$. The operator $V^0$ is the current operator \textit{in the absence of driving}. Now, we would like to know what is the current induced in the solid in response to the optical driving field.

As an illustration, we here model the solid as a system of non-interacting spinless electrons with two energy bands. The many-body Hamiltonian can be written in terms of creation/annihilation operators $c_{\mathbf{k},n}^\dag, c_{\mathbf{k},n}$ that creates/annihilates an electron in Bloch states $\ket{\mathbf{k},n}$ with quasi-momentum $\mathbf{k}$ and band $n$:
\beq
    H^0 =\sum_{\mathbf{k}}\sum_{m,n=1}^2 c_{\mathbf{k},m}^\dag h^0_{mn}(\mathbf{k})c_{\mathbf{k},n},
\eeq
where $h^0_{mn}(\mathbf{k})\equiv\bra{\mathbf{k},m}h^0\ket{\mathbf{k},n}$ is the matrix element of the first-quantized Hamiltonian $h^0$ between two Bloch states. The Bloch state can be related to the Bloch function state $\ket{u_{\mathbf{k},n}}$ via $\ket{\mathbf{k},n} = e^{i\mathbf{k}\cdot\hat{\mathbf{r}}}\ket{u_{\mathbf{k},n}}$, where $\hat{\mathbf{r}}$ is the position operator. Then we see that $h^0_{mn}(\mathbf{k}) = \bra{u_{\mathbf{k},m}}e^{-i\mathbf{k}\cdot\hat{\mathbf{r}}}h^0e^{i\mathbf{k}\cdot\hat{\mathbf{r}}}\ket{u_{\mathbf{k},n}} = \bra{u_{\mathbf{k},m}}h^0(\mathbf{k})\ket{u_{\mathbf{k},n}}$, where $h^0(\mathbf{k}) \equiv e^{-i\mathbf{k}\cdot\hat{\mathbf{r}}}h^0e^{i\mathbf{k}\cdot\hat{\mathbf{r}}}$ is a first-quantized operator.

With the equation of motion for the first-quantized position operator $\hat{\mathbf{r}}$, one can derive that the second-quantized current operator (in the absence of driving) takes the form
\beq
    V^{0} = \sum_{\mathbf{k},m,n}c_{\mathbf{k},m}^\dag \bra{u_{\mathbf{k},m}}\frac{\partial h^0(\mathbf{k})}{\partial \mathbf{k}}\ket{u_{\mathbf{k},n}}c_{\mathbf{k},n}\equiv\sum_{\mathbf{k},m,n}c_{\mathbf{k},m}^\dag v^0_{mn}(\mathbf{k}) c_{\mathbf{k},n},
\eeq
where the derivative should be understood as a directional derivative in the direction of interest. We define $\frac{\partial h^0(\mathbf{k})}{\partial\mathbf{k}}\equiv v^0(\mathbf{k})$, which is a first-quantized operator. Similarly, one can derive the time-dependent current operator in the presence of the time-dependent driving field as follows:
\beq
\begin{aligned}
    V(t) &= \sum_{\mathbf{k},m,n}c_{\mathbf{k},m}^\dag \left(\frac{\partial h(\mathbf{k},t)}{\partial \mathbf{k}}\right)_{mn}c_{\mathbf{k},n}\\
    &= \sum_{\mathbf{k},m,n}c_{\mathbf{k},m}^\dag \left\langle  u_{\mathbf{k},m} \left|\frac{\partial}{\partial \mathbf{k}}\left[h^0(\mathbf{k}) + A(t)v^0(\mathbf{k})\right]\right|u_{\mathbf{k},n}\right\rangle c_{\mathbf{k},n}\\
    &= \sum_{\mathbf{k},m,n}c_{\mathbf{k},m}^\dag \left[v^0_{mn}(\mathbf{k}) + A(t)\left(\frac{\partial v^0(\mathbf{k})}{\partial\mathbf{k}}\right)_{mn}\right]c_{\mathbf{k},n}\\
    &\equiv\sum_{\mathbf{k},m,n}c_{\mathbf{k},m}^\dag v_{mn}(\mathbf{k},t) c_{\mathbf{k},n}
\end{aligned}
\eeq

Now, it is clear that different momentum sectors do not couple with one another in this simplified model. In the situation where the band $n=1$ is the fully occupied valence band with dispersion $\epsilon_1^0(\mathbf{k})$ and $n=2$ is the empty conduction band with dispersion $\epsilon_2^0(\mathbf{k})$, each momentum sector is only occupied by one electron. Therefore, we can treat the whole system as a bunch of independent two-level systems at different momenta $\mathbf{k}$, whose time-dependent Hamiltonians (written as 2-by-2 matrices in the Bloch basis) are
\beq
    \bm{h}(\mathbf{k},t) = \bpm
        \epsilon_1^0(\mathbf{k}) + A(t)v^0_{11}(\mathbf{k})  &  A(t)v^0_{12}(\mathbf{k})\\
         A(t)v^0_{21}(\mathbf{k}) & \epsilon_2^0(\mathbf{k}) + A(t)v^0_{22}(\mathbf{k})
    \epm.
\eeq

We now wish to add dissipation to the system (in order to model the effect of a photonic heat bath), obtain the NESS $\rho_{ss}(t)$, and evaluate the response current $J(t) = {\rm Tr}(V(t)\rho_{ss}(t))$. To keep the separation between different momentum sectors, we add dissipation to the two-level system at each $\mathbf{k}$ via jump operators:
\beq
    \bm{L}_1 = \sigma_+ = \bpm 0&1\\0&0\epm, \ \ \  \bm{L}_2 = \sigma_- = \bpm 0&0\\1&0\epm,
\eeq
which represent the emission and absorption of one photon in the heat bath (not from the driving light field), respectively. 

For purposes of illustration, here we take the driving to be harmonic such that $A(t) = iAe^{-i\Omega t} - iA^*e^{i\Omega t}$, and then apply the RWA. Following the SF formalism described in the prior Appendix, we see that within the RWA there are only three Fourier components of $\bm{h}(\mathbf{k},t)$ that are nonzero:
\beq
\begin{aligned}
    &\bm{h}^{(0)}(\mathbf{k}) = \bpm
        \epsilon_1^0(\mathbf{k}) & 0\\
        0 & \epsilon_2^0(\mathbf{k})
    \epm,\\
    &\bm{h}^{(-1)}(\mathbf{k}) = -iA^*\bm{v}^0(\mathbf{k}),\\
    &\bm{h}^{(1)}(\mathbf{k}) = iA\bm{v}^0(\mathbf{k}),
\end{aligned}
\eeq
which directly lead to the SF Hamiltonian in RWA:
\beq
    \bm{h}_{SF}^{\rm RWA}(\mathbf{k}) = \bpm
        \epsilon_1^0(\mathbf{k}) + \Omega & -iA^*v^0_{12}(\mathbf{k})\\
        iAv^0_{21}(\mathbf{k}) & \epsilon_2^0(\mathbf{k})
    \epm.
\eeq

After solving for the eigenoperators of the Floquet-Lindblad equation, as is shown in the main text, we obtain the NESS within the RWA, which can be represented abstractly in the extended Hilbert space basis as:
\beq
    \bm{\rho}_{ss}^{\rm RWA}(\mathbf{k}) = \bm{\varrho}_{0}^{\rm RWA}(\mathbf{k})\equiv\bpm
        \rho_{ss,11}^{(0)}(\mathbf{k}) & \rho_{ss,12}^{(-1)}(\mathbf{k})\\
        \rho_{ss,21}^{(1)}(\mathbf{k}) & \rho_{ss,22}^{(0)}(\mathbf{k})
    \epm.
\eeq
This indicates that the NESS in the time domain is
\beq
    \bm{\rho}_{ss}^{\rm RWA}(\mathbf{k},t)\equiv\bpm
        \rho_{ss,11}^{(0)}(\mathbf{k}) & \rho_{ss,12}^{(-1)}(\mathbf{k})e^{i\Omega t}\\
        \rho_{ss,21}^{(1)}(\mathbf{k})e^{-i\Omega t} & \rho_{ss,22}^{(0)}(\mathbf{k})
    \epm.
\eeq

The current operator $V(t)$ in the momentum sector $\mathbf{k}$ can also now be represented by the following 2-by-2 matrix in Bloch basis:
\beq
\begin{aligned}
    &\bm{v}(\mathbf{k},t)\\
    =& \bpm
        v^0_{11}(\mathbf{k}) + A(t)\left(\frac{\partial v^0(\mathbf{k})}{\partial \mathbf{k}}\right)_{11}  & v^0_{12}(\mathbf{k}) + A(t)\left(\frac{\partial v^0(\mathbf{k})}{\partial \mathbf{k}}\right)_{12} \\
       v^0_{21}(\mathbf{k}) + A(t)\left(\frac{\partial v^0(\mathbf{k})}{\partial \mathbf{k}}\right)_{21}  & v^0_{22}(\mathbf{k}) + A(t)\left(\frac{\partial v^0(\mathbf{k})}{\partial \mathbf{k}}\right)_{22} 
    \epm.
\end{aligned}
\eeq
As a result, the current contributed by momentum sector $\mathbf{k}$ is given by $J(\mathbf{k},t) = {\rm Tr}(\bm{v}(\mathbf{k},t)\bm{\rho}_{ss}^{\rm RWA}(\mathbf{k},t))$. It is apparent that due to the multiplication of the terms with different oscillating frequencies ($\pm\Omega$ and $0$), the current has three components, oscillating with frequencies $0, \Omega, 2\Omega$. The component with frequency $\Omega$ is the linear response, which is not of interest, while the components with frequencies $0$ and $2\Omega$ are second-order responses, corresponding to nonlinear optical effects. 

The zero frequency component, also called the DC response, is given by:
\beq
\begin{aligned}
    J_{\rm DC}(\mathbf{k}) =& v^0_{11}(\mathbf{k})\rho_{ss,11}^{(0)}(\mathbf{k}) -iA^*\left(\frac{\partial v^0(\mathbf{k})}{\partial \mathbf{k}}\right)_{12}\rho_{ss,21}^{(1)}(\mathbf{k})\\
    &+ iA\left(\frac{\partial v^0(\mathbf{k})}{\partial \mathbf{k}}\right)_{21}\rho_{ss,12}^{(-1)}(\mathbf{k}) + v^0_{22}(\mathbf{k})\rho_{ss,22}^{(0)}(\mathbf{k})\\
    =& {\rm Tr}\left[\bm{v}^{\rm RWA}(\mathbf{k})\bm{\rho}_{ss}^{\rm RWA}(\mathbf{k})\right]
\end{aligned}
\eeq
where the matrix $v^{\rm RWA}(\mathbf{k})$ is a 2-by-2 matrix represented in the extended Hilbert space basis as
\beq
    \bm{v}^{\rm RWA}(\mathbf{k}) = \bpm
        v_{11}^0(\mathbf{k}) & -iA^*\left(\frac{\partial v^0(\mathbf{k})}{\partial \mathbf{k}}\right)_{12}\\
        iA \left(\frac{\partial v^0(\mathbf{k})}{\partial \mathbf{k}}\right)_{21} & v_{22}^0(\mathbf{k}).
    \epm
\eeq

The component with frequency $2\Omega$ is called the ``second harmonic generation (SHG)'', which is given by
\beq
\begin{aligned}
    J_{\rm SHG}(\mathbf{k},t) =& iA\left(\frac{\partial v^0(\mathbf{k})}{\partial\mathbf{k}}\right)_{12}\rho_{ss,21}^{(1)}(\mathbf{k})e^{-2i\Omega t}\\
    &- iA^*\left(\frac{\partial v^0(\mathbf{k})}{\partial \mathbf{k}}\right)_{21}\rho_{ss,12}^{(-1)}(\mathbf{k})e^{2i\Omega t} \, .
\end{aligned}
\eeq
The total current $J(t)$ is then the summation of the contributions from all momentum sectors.

Before ending the discussion of nonlinear optical effects, we would like to mention that the above results are derived under the assumption that the time-dependent Hamiltonian of the system takes the exact form of \eqref{eq:nonlinear_hamiltonian}. However, in some physical situations, especially when the light field $A(t)$ is strong, we will need to include higher-order terms of $A(t)$ in the Hamiltonian to get more precise results. This way, the time-dependent Hamiltonian will no longer be monochromatic and we will need to include more Fourier components of the NESS solution $\rho_{ss}(t)$ to obtain the nonlinear optical responses.



\bibliography{spectrum}

\begin{thebibliography}{51}%
\makeatletter
\providecommand \@ifxundefined [1]{%
 \@ifx{#1\undefined}
}%
\providecommand \@ifnum [1]{%
 \ifnum #1\expandafter \@firstoftwo
 \else \expandafter \@secondoftwo
 \fi
}%
\providecommand \@ifx [1]{%
 \ifx #1\expandafter \@firstoftwo
 \else \expandafter \@secondoftwo
 \fi
}%
\providecommand \natexlab [1]{#1}%
\providecommand \enquote  [1]{``#1''}%
\providecommand \bibnamefont  [1]{#1}%
\providecommand \bibfnamefont [1]{#1}%
\providecommand \citenamefont [1]{#1}%
\providecommand \href@noop [0]{\@secondoftwo}%
\providecommand \href [0]{\begingroup \@sanitize@url \@href}%
\providecommand \@href[1]{\@@startlink{#1}\@@href}%
\providecommand \@@href[1]{\endgroup#1\@@endlink}%
\providecommand \@sanitize@url [0]{\catcode `\\12\catcode `\$12\catcode
  `\&12\catcode `\#12\catcode `\^12\catcode `\_12\catcode `\%12\relax}%
\providecommand \@@startlink[1]{}%
\providecommand \@@endlink[0]{}%
\providecommand \url  [0]{\begingroup\@sanitize@url \@url }%
\providecommand \@url [1]{\endgroup\@href {#1}{\urlprefix }}%
\providecommand \urlprefix  [0]{URL }%
\providecommand \Eprint [0]{\href }%
\providecommand \doibase [0]{https://doi.org/}%
\providecommand \selectlanguage [0]{\@gobble}%
\providecommand \bibinfo  [0]{\@secondoftwo}%
\providecommand \bibfield  [0]{\@secondoftwo}%
\providecommand \translation [1]{[#1]}%
\providecommand \BibitemOpen [0]{}%
\providecommand \bibitemStop [0]{}%
\providecommand \bibitemNoStop [0]{.\EOS\space}%
\providecommand \EOS [0]{\spacefactor3000\relax}%
\providecommand \BibitemShut  [1]{\csname bibitem#1\endcsname}%
\let\auto@bib@innerbib\@empty
\bibitem [{\citenamefont {Preskill}(2018)}]{preskill2018quantum}%
  \BibitemOpen
  \bibfield  {author} {\bibinfo {author} {\bibfnamefont {J.}~\bibnamefont
  {Preskill}},\ }\bibfield  {title} {\bibinfo {title} {Quantum computing in the
  nisq era and beyond},\ }\href@noop {} {\bibfield  {journal} {\bibinfo
  {journal} {Quantum}\ }\textbf {\bibinfo {volume} {2}},\ \bibinfo {pages} {79}
  (\bibinfo {year} {2018})}\BibitemShut {NoStop}%
\bibitem [{\citenamefont {{Lidar}}(2019)}]{lidarreview}%
  \BibitemOpen
  \bibfield  {author} {\bibinfo {author} {\bibfnamefont {D.~A.}\ \bibnamefont
  {{Lidar}}},\ }\bibfield  {title} {\bibinfo {title} {{Lecture Notes on the
  Theory of Open Quantum Systems}},\ }\href
  {https://doi.org/10.48550/arXiv.1902.00967} {\bibfield  {journal} {\bibinfo
  {journal} {arXiv e-prints}\ ,\ \bibinfo {eid} {arXiv:1902.00967}} (\bibinfo
  {year} {2019})}\BibitemShut {NoStop}%
\bibitem [{\citenamefont {Weimer}\ \emph {et~al.}(2021)\citenamefont {Weimer},
  \citenamefont {Kshetrimayum},\ and\ \citenamefont {Or\'us}}]{weimerreview}%
  \BibitemOpen
  \bibfield  {author} {\bibinfo {author} {\bibfnamefont {H.}~\bibnamefont
  {Weimer}}, \bibinfo {author} {\bibfnamefont {A.}~\bibnamefont
  {Kshetrimayum}},\ and\ \bibinfo {author} {\bibfnamefont {R.}~\bibnamefont
  {Or\'us}},\ }\bibfield  {title} {\bibinfo {title} {Simulation methods for
  open quantum many-body systems},\ }\href
  {https://doi.org/10.1103/RevModPhys.93.015008} {\bibfield  {journal}
  {\bibinfo  {journal} {Rev. Mod. Phys.}\ }\textbf {\bibinfo {volume} {93}},\
  \bibinfo {pages} {015008} (\bibinfo {year} {2021})}\BibitemShut {NoStop}%
\bibitem [{\citenamefont {Breuer}\ and\ \citenamefont
  {Petruccione}(2002)}]{oxford}%
  \BibitemOpen
  \bibfield  {author} {\bibinfo {author} {\bibfnamefont {H.-P.}\ \bibnamefont
  {Breuer}}\ and\ \bibinfo {author} {\bibfnamefont {F.}~\bibnamefont
  {Petruccione}},\ }\href@noop {} {\emph {\bibinfo {title} {The theory of open
  quantum systems}}}\ (\bibinfo  {publisher} {Oxford University Press, USA},\
  \bibinfo {year} {2002})\BibitemShut {NoStop}%
\bibitem [{\citenamefont {Lindblad}(1976)}]{lindblad1976}%
  \BibitemOpen
  \bibfield  {author} {\bibinfo {author} {\bibfnamefont {G.}~\bibnamefont
  {Lindblad}},\ }\bibfield  {title} {\bibinfo {title} {On the generators of
  quantum dynamical semigroups},\ }\href@noop {} {\bibfield  {journal}
  {\bibinfo  {journal} {Communications in Mathematical Physics}\ }\textbf
  {\bibinfo {volume} {48}},\ \bibinfo {pages} {119} (\bibinfo {year}
  {1976})}\BibitemShut {NoStop}%
\bibitem [{\citenamefont {Gorini}\ \emph {et~al.}(1976)\citenamefont {Gorini},
  \citenamefont {Kossakowski},\ and\ \citenamefont
  {Sudarshan}}]{gorini1976completely}%
  \BibitemOpen
  \bibfield  {author} {\bibinfo {author} {\bibfnamefont {V.}~\bibnamefont
  {Gorini}}, \bibinfo {author} {\bibfnamefont {A.}~\bibnamefont
  {Kossakowski}},\ and\ \bibinfo {author} {\bibfnamefont {E.~C.~G.}\
  \bibnamefont {Sudarshan}},\ }\bibfield  {title} {\bibinfo {title} {Completely
  positive dynamical semigroups of n-level systems},\ }\href@noop {} {\bibfield
   {journal} {\bibinfo  {journal} {Journal of Mathematical Physics}\ }\textbf
  {\bibinfo {volume} {17}},\ \bibinfo {pages} {821} (\bibinfo {year}
  {1976})}\BibitemShut {NoStop}%
\bibitem [{\citenamefont {Mohseni}\ \emph {et~al.}(2008)\citenamefont
  {Mohseni}, \citenamefont {Rebentrost}, \citenamefont {Lloyd},\ and\
  \citenamefont {Aspuru-Guzik}}]{app1}%
  \BibitemOpen
  \bibfield  {author} {\bibinfo {author} {\bibfnamefont {M.}~\bibnamefont
  {Mohseni}}, \bibinfo {author} {\bibfnamefont {P.}~\bibnamefont {Rebentrost}},
  \bibinfo {author} {\bibfnamefont {S.}~\bibnamefont {Lloyd}},\ and\ \bibinfo
  {author} {\bibfnamefont {A.}~\bibnamefont {Aspuru-Guzik}},\ }\bibfield
  {title} {\bibinfo {title} {Environment-assisted quantum walks in
  photosynthetic energy transfer},\ }\href {https://doi.org/10.1063/1.3002335}
  {\bibfield  {journal} {\bibinfo  {journal} {The Journal of Chemical Physics}\
  }\textbf {\bibinfo {volume} {129}},\ \bibinfo {pages} {174106} (\bibinfo
  {year} {2008})}\BibitemShut {NoStop}%
\bibitem [{\citenamefont {Plenio}\ and\ \citenamefont {Huelga}(2008)}]{app2}%
  \BibitemOpen
  \bibfield  {author} {\bibinfo {author} {\bibfnamefont {M.~B.}\ \bibnamefont
  {Plenio}}\ and\ \bibinfo {author} {\bibfnamefont {S.~F.}\ \bibnamefont
  {Huelga}},\ }\bibfield  {title} {\bibinfo {title} {Dephasing-assisted
  transport: quantum networks and biomolecules},\ }\href
  {https://doi.org/10.1088/1367-2630/10/11/113019} {\bibfield  {journal}
  {\bibinfo  {journal} {New Journal of Physics}\ }\textbf {\bibinfo {volume}
  {10}},\ \bibinfo {pages} {113019} (\bibinfo {year} {2008})}\BibitemShut
  {NoStop}%
\bibitem [{\citenamefont {Brun}(2000)}]{app3}%
  \BibitemOpen
  \bibfield  {author} {\bibinfo {author} {\bibfnamefont {T.~A.}\ \bibnamefont
  {Brun}},\ }\bibfield  {title} {\bibinfo {title} {Continuous measurements,
  quantum trajectories, and decoherent histories},\ }\href
  {https://doi.org/10.1103/PhysRevA.61.042107} {\bibfield  {journal} {\bibinfo
  {journal} {Physical Review A}\ }\textbf {\bibinfo {volume} {61}},\ \bibinfo
  {pages} {042107} (\bibinfo {year} {2000})}\BibitemShut {NoStop}%
\bibitem [{\citenamefont {Kraus}\ \emph {et~al.}(2008)\citenamefont {Kraus},
  \citenamefont {Büchler}, \citenamefont {Diehl}, \citenamefont {Kantian},
  \citenamefont {Micheli},\ and\ \citenamefont {Zoller}}]{app4}%
  \BibitemOpen
  \bibfield  {author} {\bibinfo {author} {\bibfnamefont {B.}~\bibnamefont
  {Kraus}}, \bibinfo {author} {\bibfnamefont {H.~P.}\ \bibnamefont {Büchler}},
  \bibinfo {author} {\bibfnamefont {S.}~\bibnamefont {Diehl}}, \bibinfo
  {author} {\bibfnamefont {A.}~\bibnamefont {Kantian}}, \bibinfo {author}
  {\bibfnamefont {A.}~\bibnamefont {Micheli}},\ and\ \bibinfo {author}
  {\bibfnamefont {P.}~\bibnamefont {Zoller}},\ }\bibfield  {title} {\bibinfo
  {title} {Preparation of entangled states by quantum {Markov} processes},\
  }\href {https://doi.org/10.1103/PhysRevA.78.042307} {\bibfield  {journal}
  {\bibinfo  {journal} {Physical Review A}\ }\textbf {\bibinfo {volume} {78}},\
  \bibinfo {pages} {042307} (\bibinfo {year} {2008})}\BibitemShut {NoStop}%
\bibitem [{\citenamefont {Lidar}\ \emph {et~al.}(1998)\citenamefont {Lidar},
  \citenamefont {Chuang},\ and\ \citenamefont {Whaley}}]{app5}%
  \BibitemOpen
  \bibfield  {author} {\bibinfo {author} {\bibfnamefont {D.~A.}\ \bibnamefont
  {Lidar}}, \bibinfo {author} {\bibfnamefont {I.~L.}\ \bibnamefont {Chuang}},\
  and\ \bibinfo {author} {\bibfnamefont {K.~B.}\ \bibnamefont {Whaley}},\
  }\bibfield  {title} {\bibinfo {title} {Decoherence-{Free} {Subspaces} for
  {Quantum} {Computation}},\ }\href
  {https://doi.org/10.1103/PhysRevLett.81.2594} {\bibfield  {journal} {\bibinfo
   {journal} {Physical Review Letters}\ }\textbf {\bibinfo {volume} {81}},\
  \bibinfo {pages} {2594} (\bibinfo {year} {1998})}\BibitemShut {NoStop}%
\bibitem [{\citenamefont {Jones}\ \emph {et~al.}(2018)\citenamefont {Jones},
  \citenamefont {Needham}, \citenamefont {Lesanovsky}, \citenamefont
  {Intravaia},\ and\ \citenamefont {Olmos}}]{app6}%
  \BibitemOpen
  \bibfield  {author} {\bibinfo {author} {\bibfnamefont {R.}~\bibnamefont
  {Jones}}, \bibinfo {author} {\bibfnamefont {J.~A.}\ \bibnamefont {Needham}},
  \bibinfo {author} {\bibfnamefont {I.}~\bibnamefont {Lesanovsky}}, \bibinfo
  {author} {\bibfnamefont {F.}~\bibnamefont {Intravaia}},\ and\ \bibinfo
  {author} {\bibfnamefont {B.}~\bibnamefont {Olmos}},\ }\bibfield  {title}
  {\bibinfo {title} {Modified dipole-dipole interaction and dissipation in an
  atomic ensemble near surfaces},\ }\href
  {https://doi.org/10.1103/PhysRevA.97.053841} {\bibfield  {journal} {\bibinfo
  {journal} {Physical Review A}\ }\textbf {\bibinfo {volume} {97}},\ \bibinfo
  {pages} {053841} (\bibinfo {year} {2018})}\BibitemShut {NoStop}%
\bibitem [{\citenamefont {Metz}\ \emph {et~al.}(2006)\citenamefont {Metz},
  \citenamefont {Trupke},\ and\ \citenamefont {Beige}}]{app7}%
  \BibitemOpen
  \bibfield  {author} {\bibinfo {author} {\bibfnamefont {J.}~\bibnamefont
  {Metz}}, \bibinfo {author} {\bibfnamefont {M.}~\bibnamefont {Trupke}},\ and\
  \bibinfo {author} {\bibfnamefont {A.}~\bibnamefont {Beige}},\ }\bibfield
  {title} {\bibinfo {title} {Robust {Entanglement} through {Macroscopic}
  {Quantum} {Jumps}},\ }\href {https://doi.org/10.1103/PhysRevLett.97.040503}
  {\bibfield  {journal} {\bibinfo  {journal} {Physical Review Letters}\
  }\textbf {\bibinfo {volume} {97}},\ \bibinfo {pages} {040503} (\bibinfo
  {year} {2006})}\BibitemShut {NoStop}%
\bibitem [{\citenamefont {Olmos}\ \emph {et~al.}(2012)\citenamefont {Olmos},
  \citenamefont {Lesanovsky},\ and\ \citenamefont {Garrahan}}]{app8}%
  \BibitemOpen
  \bibfield  {author} {\bibinfo {author} {\bibfnamefont {B.}~\bibnamefont
  {Olmos}}, \bibinfo {author} {\bibfnamefont {I.}~\bibnamefont {Lesanovsky}},\
  and\ \bibinfo {author} {\bibfnamefont {J.~P.}\ \bibnamefont {Garrahan}},\
  }\bibfield  {title} {\bibinfo {title} {Facilitated {Spin} {Models} of
  {Dissipative} {Quantum} {Glasses}},\ }\href
  {https://doi.org/10.1103/PhysRevLett.109.020403} {\bibfield  {journal}
  {\bibinfo  {journal} {Physical Review Letters}\ }\textbf {\bibinfo {volume}
  {109}},\ \bibinfo {pages} {020403} (\bibinfo {year} {2012})}\BibitemShut
  {NoStop}%
\bibitem [{\citenamefont {Manzano}\ \emph {et~al.}(2016)\citenamefont
  {Manzano}, \citenamefont {Chuang},\ and\ \citenamefont {Cao}}]{app9}%
  \BibitemOpen
  \bibfield  {author} {\bibinfo {author} {\bibfnamefont {D.}~\bibnamefont
  {Manzano}}, \bibinfo {author} {\bibfnamefont {C.}~\bibnamefont {Chuang}},\
  and\ \bibinfo {author} {\bibfnamefont {J.}~\bibnamefont {Cao}},\ }\bibfield
  {title} {\bibinfo {title} {Quantum transport in \textit{d} -dimensional
  lattices},\ }\href {https://doi.org/10.1088/1367-2630/18/4/043044} {\bibfield
   {journal} {\bibinfo  {journal} {New Journal of Physics}\ }\textbf {\bibinfo
  {volume} {18}},\ \bibinfo {pages} {043044} (\bibinfo {year}
  {2016})}\BibitemShut {NoStop}%
\bibitem [{\citenamefont {Manzano}\ \emph {et~al.}(2012)\citenamefont
  {Manzano}, \citenamefont {Tiersch}, \citenamefont {Asadian},\ and\
  \citenamefont {Briegel}}]{app10}%
  \BibitemOpen
  \bibfield  {author} {\bibinfo {author} {\bibfnamefont {D.}~\bibnamefont
  {Manzano}}, \bibinfo {author} {\bibfnamefont {M.}~\bibnamefont {Tiersch}},
  \bibinfo {author} {\bibfnamefont {A.}~\bibnamefont {Asadian}},\ and\ \bibinfo
  {author} {\bibfnamefont {H.~J.}\ \bibnamefont {Briegel}},\ }\bibfield
  {title} {\bibinfo {title} {Quantum transport efficiency and {Fourier}'s
  law},\ }\href {https://doi.org/10.1103/PhysRevE.86.061118} {\bibfield
  {journal} {\bibinfo  {journal} {Physical Review E}\ }\textbf {\bibinfo
  {volume} {86}},\ \bibinfo {pages} {061118} (\bibinfo {year}
  {2012})}\BibitemShut {NoStop}%
\bibitem [{\citenamefont {Prosen}(2011)}]{app11}%
  \BibitemOpen
  \bibfield  {author} {\bibinfo {author} {\bibfnamefont {T.}~\bibnamefont
  {Prosen}},\ }\bibfield  {title} {\bibinfo {title} {Open {X} {X} {Z} {Spin}
  {Chain}: {Nonequilibrium} {Steady} {State} and a {Strict} {Bound} on
  {Ballistic} {Transport}},\ }\href
  {https://doi.org/10.1103/PhysRevLett.106.217206} {\bibfield  {journal}
  {\bibinfo  {journal} {Physical Review Letters}\ }\textbf {\bibinfo {volume}
  {106}},\ \bibinfo {pages} {217206} (\bibinfo {year} {2011})}\BibitemShut
  {NoStop}%
\bibitem [{\citenamefont {Manzano}\ and\ \citenamefont
  {Kyoseva}(2016)}]{app12}%
  \BibitemOpen
  \bibfield  {author} {\bibinfo {author} {\bibfnamefont {D.}~\bibnamefont
  {Manzano}}\ and\ \bibinfo {author} {\bibfnamefont {E.}~\bibnamefont
  {Kyoseva}},\ }\bibfield  {title} {\bibinfo {title} {An atomic
  symmetry-controlled thermal switch},\ }\href
  {https://doi.org/10.1038/srep31161} {\bibfield  {journal} {\bibinfo
  {journal} {Scientific Reports}\ }\textbf {\bibinfo {volume} {6}},\ \bibinfo
  {pages} {31161} (\bibinfo {year} {2016})}\BibitemShut {NoStop}%
\bibitem [{\citenamefont {Tupkary}\ \emph {et~al.}(2022)\citenamefont
  {Tupkary}, \citenamefont {Dhar}, \citenamefont {Kulkarni},\ and\
  \citenamefont {Purkayastha}}]{dharlimits}%
  \BibitemOpen
  \bibfield  {author} {\bibinfo {author} {\bibfnamefont {D.}~\bibnamefont
  {Tupkary}}, \bibinfo {author} {\bibfnamefont {A.}~\bibnamefont {Dhar}},
  \bibinfo {author} {\bibfnamefont {M.}~\bibnamefont {Kulkarni}},\ and\
  \bibinfo {author} {\bibfnamefont {A.}~\bibnamefont {Purkayastha}},\
  }\bibfield  {title} {\bibinfo {title} {Fundamental limitations in lindblad
  descriptions of systems weakly coupled to baths},\ }\href
  {https://doi.org/10.1103/PhysRevA.105.032208} {\bibfield  {journal} {\bibinfo
   {journal} {Phys. Rev. A}\ }\textbf {\bibinfo {volume} {105}},\ \bibinfo
  {pages} {032208} (\bibinfo {year} {2022})}\BibitemShut {NoStop}%
\bibitem [{\citenamefont {Manzano}(2020)}]{AJP}%
  \BibitemOpen
  \bibfield  {author} {\bibinfo {author} {\bibfnamefont {D.}~\bibnamefont
  {Manzano}},\ }\bibfield  {title} {\bibinfo {title} {A short introduction to
  the lindblad master equation},\ }\href@noop {} {\bibfield  {journal}
  {\bibinfo  {journal} {Aip Advances}\ }\textbf {\bibinfo {volume} {10}}
  (\bibinfo {year} {2020})}\BibitemShut {NoStop}%
\bibitem [{Note1()}]{Note1}%
  \BibitemOpen
  \bibinfo {note} {The condition of the dynamics being ``completely'' positive
  is more general than simply keeping a non-negative $\rho (t_0)$ non-negative
  at future times $t>t_0$, which is the positivity condition. In other words,
  not all positive dynamics are completely positive: see Ref.~\cite {AJP} for
  more details.}\BibitemShut {Stop}%
\bibitem [{\citenamefont {Barnett}\ and\ \citenamefont
  {Stenholm}(2000)}]{spectral_2000}%
  \BibitemOpen
  \bibfield  {author} {\bibinfo {author} {\bibfnamefont {S.~M.}\ \bibnamefont
  {Barnett}}\ and\ \bibinfo {author} {\bibfnamefont {S.}~\bibnamefont
  {Stenholm}},\ }\bibfield  {title} {\bibinfo {title} {Spectral decomposition
  of the lindblad operator},\ }\href@noop {} {\bibfield  {journal} {\bibinfo
  {journal} {Journal of Modern Optics}\ }\textbf {\bibinfo {volume} {47}},\
  \bibinfo {pages} {2869} (\bibinfo {year} {2000})}\BibitemShut {NoStop}%
\bibitem [{\citenamefont {Albert}\ and\ \citenamefont
  {Jiang}(2014)}]{spectrum_Jiang}%
  \BibitemOpen
  \bibfield  {author} {\bibinfo {author} {\bibfnamefont {V.~V.}\ \bibnamefont
  {Albert}}\ and\ \bibinfo {author} {\bibfnamefont {L.}~\bibnamefont {Jiang}},\
  }\bibfield  {title} {\bibinfo {title} {Symmetries and conserved quantities in
  lindblad master equations},\ }\href@noop {} {\bibfield  {journal} {\bibinfo
  {journal} {Physical Review A}\ }\textbf {\bibinfo {volume} {89}},\ \bibinfo
  {pages} {022118} (\bibinfo {year} {2014})}\BibitemShut {NoStop}%
\bibitem [{\citenamefont {Minganti}\ \emph {et~al.}(2018)\citenamefont
  {Minganti}, \citenamefont {Biella}, \citenamefont {Bartolo},\ and\
  \citenamefont {Ciuti}}]{spectrum_proof}%
  \BibitemOpen
  \bibfield  {author} {\bibinfo {author} {\bibfnamefont {F.}~\bibnamefont
  {Minganti}}, \bibinfo {author} {\bibfnamefont {A.}~\bibnamefont {Biella}},
  \bibinfo {author} {\bibfnamefont {N.}~\bibnamefont {Bartolo}},\ and\ \bibinfo
  {author} {\bibfnamefont {C.}~\bibnamefont {Ciuti}},\ }\bibfield  {title}
  {\bibinfo {title} {Spectral theory of liouvillians for dissipative phase
  transitions},\ }\href@noop {} {\bibfield  {journal} {\bibinfo  {journal}
  {Physical Review A}\ }\textbf {\bibinfo {volume} {98}},\ \bibinfo {pages}
  {042118} (\bibinfo {year} {2018})}\BibitemShut {NoStop}%
\bibitem [{\citenamefont {Minganti}\ and\ \citenamefont
  {Huybrechts}(2022)}]{spectrum_Floquet}%
  \BibitemOpen
  \bibfield  {author} {\bibinfo {author} {\bibfnamefont {F.}~\bibnamefont
  {Minganti}}\ and\ \bibinfo {author} {\bibfnamefont {D.}~\bibnamefont
  {Huybrechts}},\ }\bibfield  {title} {\bibinfo {title} {Arnoldi-lindblad time
  evolution: Faster-than-the-clock algorithm for the spectrum of
  time-independent and floquet open quantum systems},\ }\href@noop {}
  {\bibfield  {journal} {\bibinfo  {journal} {Quantum}\ }\textbf {\bibinfo
  {volume} {6}},\ \bibinfo {pages} {649} (\bibinfo {year} {2022})}\BibitemShut
  {NoStop}%
\bibitem [{\citenamefont {Schmiedinghoff}\ and\ \citenamefont
  {Uhrig}(2022)}]{Gary_contact}%
  \BibitemOpen
  \bibfield  {author} {\bibinfo {author} {\bibfnamefont {G.}~\bibnamefont
  {Schmiedinghoff}}\ and\ \bibinfo {author} {\bibfnamefont {G.~S.}\
  \bibnamefont {Uhrig}},\ }\bibfield  {title} {\bibinfo {title} {{Efficient
  flow equations for dissipative systems}},\ }\href
  {https://doi.org/10.21468/SciPostPhys.13.6.122} {\bibfield  {journal}
  {\bibinfo  {journal} {SciPost Phys.}\ }\textbf {\bibinfo {volume} {13}},\
  \bibinfo {pages} {122} (\bibinfo {year} {2022})}\BibitemShut {NoStop}%
\bibitem [{\citenamefont {Fitzpatrick}\ \emph {et~al.}(2017)\citenamefont
  {Fitzpatrick}, \citenamefont {Sundaresan}, \citenamefont {Li}, \citenamefont
  {Koch},\ and\ \citenamefont {Houck}}]{fitzpatrick2017}%
  \BibitemOpen
  \bibfield  {author} {\bibinfo {author} {\bibfnamefont {M.}~\bibnamefont
  {Fitzpatrick}}, \bibinfo {author} {\bibfnamefont {N.~M.}\ \bibnamefont
  {Sundaresan}}, \bibinfo {author} {\bibfnamefont {A.~C.~Y.}\ \bibnamefont
  {Li}}, \bibinfo {author} {\bibfnamefont {J.}~\bibnamefont {Koch}},\ and\
  \bibinfo {author} {\bibfnamefont {A.~A.}\ \bibnamefont {Houck}},\ }\bibfield
  {title} {\bibinfo {title} {Observation of a dissipative phase transition in a
  one-dimensional circuit qed lattice},\ }\href
  {https://doi.org/10.1103/PhysRevX.7.011016} {\bibfield  {journal} {\bibinfo
  {journal} {Phys. Rev. X}\ }\textbf {\bibinfo {volume} {7}},\ \bibinfo {pages}
  {011016} (\bibinfo {year} {2017})}\BibitemShut {NoStop}%
\bibitem [{\citenamefont {Fedorov}\ \emph {et~al.}(2021)\citenamefont
  {Fedorov}, \citenamefont {Remizov}, \citenamefont {Shapiro}, \citenamefont
  {Pogosov}, \citenamefont {Egorova}, \citenamefont {Tsitsilin}, \citenamefont
  {Andronik}, \citenamefont {Dobronosova}, \citenamefont {Rodionov},
  \citenamefont {Astafiev},\ and\ \citenamefont {Ustinov}}]{fedorov2021}%
  \BibitemOpen
  \bibfield  {author} {\bibinfo {author} {\bibfnamefont {G.~P.}\ \bibnamefont
  {Fedorov}}, \bibinfo {author} {\bibfnamefont {S.~V.}\ \bibnamefont
  {Remizov}}, \bibinfo {author} {\bibfnamefont {D.~S.}\ \bibnamefont
  {Shapiro}}, \bibinfo {author} {\bibfnamefont {W.~V.}\ \bibnamefont
  {Pogosov}}, \bibinfo {author} {\bibfnamefont {E.}~\bibnamefont {Egorova}},
  \bibinfo {author} {\bibfnamefont {I.}~\bibnamefont {Tsitsilin}}, \bibinfo
  {author} {\bibfnamefont {M.}~\bibnamefont {Andronik}}, \bibinfo {author}
  {\bibfnamefont {A.~A.}\ \bibnamefont {Dobronosova}}, \bibinfo {author}
  {\bibfnamefont {I.~A.}\ \bibnamefont {Rodionov}}, \bibinfo {author}
  {\bibfnamefont {O.~V.}\ \bibnamefont {Astafiev}},\ and\ \bibinfo {author}
  {\bibfnamefont {A.~V.}\ \bibnamefont {Ustinov}},\ }\bibfield  {title}
  {\bibinfo {title} {Photon transport in a bose-hubbard chain of
  superconducting artificial atoms},\ }\href
  {https://doi.org/10.1103/PhysRevLett.126.180503} {\bibfield  {journal}
  {\bibinfo  {journal} {Phys. Rev. Lett.}\ }\textbf {\bibinfo {volume} {126}},\
  \bibinfo {pages} {180503} (\bibinfo {year} {2021})}\BibitemShut {NoStop}%
\bibitem [{\citenamefont {Mori}(2023)}]{open_floquet_review}%
  \BibitemOpen
  \bibfield  {author} {\bibinfo {author} {\bibfnamefont {T.}~\bibnamefont
  {Mori}},\ }\bibfield  {title} {\bibinfo {title} {Floquet states in open
  quantum systems},\ }\href@noop {} {\bibfield  {journal} {\bibinfo  {journal}
  {Annual Review of Condensed Matter Physics}\ }\textbf {\bibinfo {volume}
  {14}},\ \bibinfo {pages} {35} (\bibinfo {year} {2023})}\BibitemShut {NoStop}%
\bibitem [{\citenamefont {Haddadfarshi}\ \emph {et~al.}(2015)\citenamefont
  {Haddadfarshi}, \citenamefont {Cui},\ and\ \citenamefont
  {Mintert}}]{open_floquet1}%
  \BibitemOpen
  \bibfield  {author} {\bibinfo {author} {\bibfnamefont {F.}~\bibnamefont
  {Haddadfarshi}}, \bibinfo {author} {\bibfnamefont {J.}~\bibnamefont {Cui}},\
  and\ \bibinfo {author} {\bibfnamefont {F.}~\bibnamefont {Mintert}},\
  }\bibfield  {title} {\bibinfo {title} {Completely positive approximate
  solutions of driven open quantum systems},\ }\href@noop {} {\bibfield
  {journal} {\bibinfo  {journal} {Physical Review Letters}\ }\textbf {\bibinfo
  {volume} {114}},\ \bibinfo {pages} {130402} (\bibinfo {year}
  {2015})}\BibitemShut {NoStop}%
\bibitem [{\citenamefont {Dai}\ \emph {et~al.}(2016)\citenamefont {Dai},
  \citenamefont {Shi},\ and\ \citenamefont {Yi}}]{open_floquet2}%
  \BibitemOpen
  \bibfield  {author} {\bibinfo {author} {\bibfnamefont {C.}~\bibnamefont
  {Dai}}, \bibinfo {author} {\bibfnamefont {Z.}~\bibnamefont {Shi}},\ and\
  \bibinfo {author} {\bibfnamefont {X.}~\bibnamefont {Yi}},\ }\bibfield
  {title} {\bibinfo {title} {Floquet theorem with open systems and its
  applications},\ }\href@noop {} {\bibfield  {journal} {\bibinfo  {journal}
  {Physical Review A}\ }\textbf {\bibinfo {volume} {93}},\ \bibinfo {pages}
  {032121} (\bibinfo {year} {2016})}\BibitemShut {NoStop}%
\bibitem [{\citenamefont {Hartmann}\ \emph {et~al.}(2017)\citenamefont
  {Hartmann}, \citenamefont {Poletti}, \citenamefont {Ivanchenko},
  \citenamefont {Denisov},\ and\ \citenamefont {H{\"a}nggi}}]{open_floquet3}%
  \BibitemOpen
  \bibfield  {author} {\bibinfo {author} {\bibfnamefont {M.}~\bibnamefont
  {Hartmann}}, \bibinfo {author} {\bibfnamefont {D.}~\bibnamefont {Poletti}},
  \bibinfo {author} {\bibfnamefont {M.}~\bibnamefont {Ivanchenko}}, \bibinfo
  {author} {\bibfnamefont {S.}~\bibnamefont {Denisov}},\ and\ \bibinfo {author}
  {\bibfnamefont {P.}~\bibnamefont {H{\"a}nggi}},\ }\bibfield  {title}
  {\bibinfo {title} {Asymptotic floquet states of open quantum systems: the
  role of interaction},\ }\href@noop {} {\bibfield  {journal} {\bibinfo
  {journal} {New Journal of Physics}\ }\textbf {\bibinfo {volume} {19}},\
  \bibinfo {pages} {083011} (\bibinfo {year} {2017})}\BibitemShut {NoStop}%
\bibitem [{\citenamefont {Schnell}\ \emph {et~al.}(2020)\citenamefont
  {Schnell}, \citenamefont {Eckardt},\ and\ \citenamefont
  {Denisov}}]{is_there}%
  \BibitemOpen
  \bibfield  {author} {\bibinfo {author} {\bibfnamefont {A.}~\bibnamefont
  {Schnell}}, \bibinfo {author} {\bibfnamefont {A.}~\bibnamefont {Eckardt}},\
  and\ \bibinfo {author} {\bibfnamefont {S.}~\bibnamefont {Denisov}},\
  }\bibfield  {title} {\bibinfo {title} {Is there a floquet lindbladian?},\
  }\href@noop {} {\bibfield  {journal} {\bibinfo  {journal} {Physical Review
  B}\ }\textbf {\bibinfo {volume} {101}},\ \bibinfo {pages} {100301} (\bibinfo
  {year} {2020})}\BibitemShut {NoStop}%
\bibitem [{\citenamefont {Holthaus}(2015)}]{floquet_notes}%
  \BibitemOpen
  \bibfield  {author} {\bibinfo {author} {\bibfnamefont {M.}~\bibnamefont
  {Holthaus}},\ }\bibfield  {title} {\bibinfo {title} {Floquet engineering with
  quasienergy bands of periodically driven optical lattices},\ }\href
  {https://doi.org/10.1088/0953-4075/49/1/013001} {\bibfield  {journal}
  {\bibinfo  {journal} {Journal of Physics B: Atomic, Molecular and Optical
  Physics}\ }\textbf {\bibinfo {volume} {49}},\ \bibinfo {pages} {013001}
  (\bibinfo {year} {2015})}\BibitemShut {NoStop}%
\bibitem [{\citenamefont {Wolf}\ \emph {et~al.}(2008)\citenamefont {Wolf},
  \citenamefont {Eisert}, \citenamefont {Cubitt},\ and\ \citenamefont
  {Cirac}}]{open_floquet_nonmarkovian}%
  \BibitemOpen
  \bibfield  {author} {\bibinfo {author} {\bibfnamefont {M.~M.}\ \bibnamefont
  {Wolf}}, \bibinfo {author} {\bibfnamefont {J.}~\bibnamefont {Eisert}},
  \bibinfo {author} {\bibfnamefont {T.~S.}\ \bibnamefont {Cubitt}},\ and\
  \bibinfo {author} {\bibfnamefont {J.~I.}\ \bibnamefont {Cirac}},\ }\bibfield
  {title} {\bibinfo {title} {Assessing non-markovian quantum dynamics},\
  }\href@noop {} {\bibfield  {journal} {\bibinfo  {journal} {Physical review
  letters}\ }\textbf {\bibinfo {volume} {101}},\ \bibinfo {pages} {150402}
  (\bibinfo {year} {2008})}\BibitemShut {NoStop}%
\bibitem [{\citenamefont {Mizuta}\ \emph {et~al.}(2021)\citenamefont {Mizuta},
  \citenamefont {Takasan},\ and\ \citenamefont
  {Kawakami}}]{floquet_lindbladian1}%
  \BibitemOpen
  \bibfield  {author} {\bibinfo {author} {\bibfnamefont {K.}~\bibnamefont
  {Mizuta}}, \bibinfo {author} {\bibfnamefont {K.}~\bibnamefont {Takasan}},\
  and\ \bibinfo {author} {\bibfnamefont {N.}~\bibnamefont {Kawakami}},\
  }\bibfield  {title} {\bibinfo {title} {Breakdown of markovianity by
  interactions in stroboscopic floquet-lindblad dynamics under high-frequency
  drive},\ }\href@noop {} {\bibfield  {journal} {\bibinfo  {journal} {Physical
  Review A}\ }\textbf {\bibinfo {volume} {103}},\ \bibinfo {pages} {L020202}
  (\bibinfo {year} {2021})}\BibitemShut {NoStop}%
\bibitem [{\citenamefont {Schnell}\ \emph {et~al.}(2021)\citenamefont
  {Schnell}, \citenamefont {Denisov},\ and\ \citenamefont
  {Eckardt}}]{floquet_lindbladian2}%
  \BibitemOpen
  \bibfield  {author} {\bibinfo {author} {\bibfnamefont {A.}~\bibnamefont
  {Schnell}}, \bibinfo {author} {\bibfnamefont {S.}~\bibnamefont {Denisov}},\
  and\ \bibinfo {author} {\bibfnamefont {A.}~\bibnamefont {Eckardt}},\
  }\bibfield  {title} {\bibinfo {title} {High-frequency expansions for
  time-periodic lindblad generators},\ }\href@noop {} {\bibfield  {journal}
  {\bibinfo  {journal} {Physical Review B}\ }\textbf {\bibinfo {volume}
  {104}},\ \bibinfo {pages} {165414} (\bibinfo {year} {2021})}\BibitemShut
  {NoStop}%
\bibitem [{\citenamefont {Ikeda}\ \emph {et~al.}(2021)\citenamefont {Ikeda},
  \citenamefont {Chinzei},\ and\ \citenamefont {Sato}}]{floquet_lindbladian3}%
  \BibitemOpen
  \bibfield  {author} {\bibinfo {author} {\bibfnamefont {T.}~\bibnamefont
  {Ikeda}}, \bibinfo {author} {\bibfnamefont {K.}~\bibnamefont {Chinzei}},\
  and\ \bibinfo {author} {\bibfnamefont {M.}~\bibnamefont {Sato}},\ }\bibfield
  {title} {\bibinfo {title} {Nonequilibrium steady states in the
  floquet-lindblad systems: van vleck's high-frequency expansion approach},\
  }\href@noop {} {\bibfield  {journal} {\bibinfo  {journal} {SciPost Physics
  Core}\ }\textbf {\bibinfo {volume} {4}},\ \bibinfo {pages} {033} (\bibinfo
  {year} {2021})}\BibitemShut {NoStop}%
\bibitem [{\citenamefont {Yudin}\ \emph {et~al.}(2016)\citenamefont {Yudin},
  \citenamefont {Taichenachev},\ and\ \citenamefont
  {Basalaev}}]{no_floquet_contact}%
  \BibitemOpen
  \bibfield  {author} {\bibinfo {author} {\bibfnamefont {V.~I.}\ \bibnamefont
  {Yudin}}, \bibinfo {author} {\bibfnamefont {A.~V.}\ \bibnamefont
  {Taichenachev}},\ and\ \bibinfo {author} {\bibfnamefont {M.~Y.}\ \bibnamefont
  {Basalaev}},\ }\bibfield  {title} {\bibinfo {title} {Dynamic steady state of
  periodically driven quantum systems},\ }\href
  {https://doi.org/10.1103/PhysRevA.93.013820} {\bibfield  {journal} {\bibinfo
  {journal} {Phys. Rev. A}\ }\textbf {\bibinfo {volume} {93}},\ \bibinfo
  {pages} {013820} (\bibinfo {year} {2016})}\BibitemShut {NoStop}%
\bibitem [{\citenamefont {Morimoto}\ and\ \citenamefont
  {Nagaosa}(2016)}]{nagaosa_2016}%
  \BibitemOpen
  \bibfield  {author} {\bibinfo {author} {\bibfnamefont {T.}~\bibnamefont
  {Morimoto}}\ and\ \bibinfo {author} {\bibfnamefont {N.}~\bibnamefont
  {Nagaosa}},\ }\bibfield  {title} {\bibinfo {title} {Topological nature of
  nonlinear optical effects in solids},\ }\href@noop {} {\bibfield  {journal}
  {\bibinfo  {journal} {Science advances}\ }\textbf {\bibinfo {volume} {2}},\
  \bibinfo {pages} {e1501524} (\bibinfo {year} {2016})}\BibitemShut {NoStop}%
\bibitem [{\citenamefont {Ho}\ \emph {et~al.}(1986)\citenamefont {Ho},
  \citenamefont {Wang},\ and\ \citenamefont {Chu}}]{Fl1}%
  \BibitemOpen
  \bibfield  {author} {\bibinfo {author} {\bibfnamefont {T.-S.}\ \bibnamefont
  {Ho}}, \bibinfo {author} {\bibfnamefont {K.}~\bibnamefont {Wang}},\ and\
  \bibinfo {author} {\bibfnamefont {S.-I.}\ \bibnamefont {Chu}},\ }\bibfield
  {title} {\bibinfo {title} {Floquet-liouville supermatrix approach: Time
  development of density-matrix operator and multiphoton resonance fluorescence
  spectra in intense laser fields},\ }\href
  {https://doi.org/10.1103/PhysRevA.33.1798} {\bibfield  {journal} {\bibinfo
  {journal} {Phys. Rev. A}\ }\textbf {\bibinfo {volume} {33}},\ \bibinfo
  {pages} {1798} (\bibinfo {year} {1986})}\BibitemShut {NoStop}%
\bibitem [{\citenamefont {Prosen}\ and\ \citenamefont {Ilievski}(2011)}]{FL2}%
  \BibitemOpen
  \bibfield  {author} {\bibinfo {author} {\bibfnamefont {T.~c.~v.}\
  \bibnamefont {Prosen}}\ and\ \bibinfo {author} {\bibfnamefont
  {E.}~\bibnamefont {Ilievski}},\ }\bibfield  {title} {\bibinfo {title}
  {Nonequilibrium phase transition in a periodically driven $xy$ spin chain},\
  }\href {https://doi.org/10.1103/PhysRevLett.107.060403} {\bibfield  {journal}
  {\bibinfo  {journal} {Phys. Rev. Lett.}\ }\textbf {\bibinfo {volume} {107}},\
  \bibinfo {pages} {060403} (\bibinfo {year} {2011})}\BibitemShut {NoStop}%
\bibitem [{\citenamefont {Vorberg}\ \emph {et~al.}(2013)\citenamefont
  {Vorberg}, \citenamefont {Wustmann}, \citenamefont {Ketzmerick},\ and\
  \citenamefont {Eckardt}}]{FL3}%
  \BibitemOpen
  \bibfield  {author} {\bibinfo {author} {\bibfnamefont {D.}~\bibnamefont
  {Vorberg}}, \bibinfo {author} {\bibfnamefont {W.}~\bibnamefont {Wustmann}},
  \bibinfo {author} {\bibfnamefont {R.}~\bibnamefont {Ketzmerick}},\ and\
  \bibinfo {author} {\bibfnamefont {A.}~\bibnamefont {Eckardt}},\ }\bibfield
  {title} {\bibinfo {title} {Generalized bose-einstein condensation into
  multiple states in driven-dissipative systems},\ }\href
  {https://doi.org/10.1103/PhysRevLett.111.240405} {\bibfield  {journal}
  {\bibinfo  {journal} {Phys. Rev. Lett.}\ }\textbf {\bibinfo {volume} {111}},\
  \bibinfo {pages} {240405} (\bibinfo {year} {2013})}\BibitemShut {NoStop}%
\bibitem [{\citenamefont {Ikeda}\ and\ \citenamefont
  {Sato}(2020)}]{FLSSgeneral}%
  \BibitemOpen
  \bibfield  {author} {\bibinfo {author} {\bibfnamefont {T.~N.}\ \bibnamefont
  {Ikeda}}\ and\ \bibinfo {author} {\bibfnamefont {M.}~\bibnamefont {Sato}},\
  }\bibfield  {title} {\bibinfo {title} {General description for nonequilibrium
  steady states in periodically driven dissipative quantum systems},\ }\href
  {https://www.science.org/doi/full/10.1126/sciadv.abb4019} {\bibfield
  {journal} {\bibinfo  {journal} {Science advances}\ }\textbf {\bibinfo
  {volume} {6}},\ \bibinfo {pages} {eabb4019} (\bibinfo {year}
  {2020})}\BibitemShut {NoStop}%
\bibitem [{\citenamefont {Chinzei}\ and\ \citenamefont
  {Ikeda}(2020)}]{timecrystal}%
  \BibitemOpen
  \bibfield  {author} {\bibinfo {author} {\bibfnamefont {K.}~\bibnamefont
  {Chinzei}}\ and\ \bibinfo {author} {\bibfnamefont {T.~N.}\ \bibnamefont
  {Ikeda}},\ }\bibfield  {title} {\bibinfo {title} {Time crystals protected by
  floquet dynamical symmetry in hubbard models},\ }\href
  {https://doi.org/10.1103/PhysRevLett.125.060601} {\bibfield  {journal}
  {\bibinfo  {journal} {Phys. Rev. Lett.}\ }\textbf {\bibinfo {volume} {125}},\
  \bibinfo {pages} {060601} (\bibinfo {year} {2020})}\BibitemShut {NoStop}%
\bibitem [{\citenamefont {Kohler}\ \emph {et~al.}(2005)\citenamefont {Kohler},
  \citenamefont {Lehmann},\ and\ \citenamefont {H{\"a}nggi}}]{keldysh1}%
  \BibitemOpen
  \bibfield  {author} {\bibinfo {author} {\bibfnamefont {S.}~\bibnamefont
  {Kohler}}, \bibinfo {author} {\bibfnamefont {J.}~\bibnamefont {Lehmann}},\
  and\ \bibinfo {author} {\bibfnamefont {P.}~\bibnamefont {H{\"a}nggi}},\
  }\bibfield  {title} {\bibinfo {title} {Driven quantum transport on the
  nanoscale},\ }\href@noop {} {\bibfield  {journal} {\bibinfo  {journal}
  {Physics Reports}\ }\textbf {\bibinfo {volume} {406}},\ \bibinfo {pages}
  {379} (\bibinfo {year} {2005})}\BibitemShut {NoStop}%
\bibitem [{\citenamefont {Jauho}\ \emph {et~al.}(1994)\citenamefont {Jauho},
  \citenamefont {Wingreen},\ and\ \citenamefont {Meir}}]{keldysh2}%
  \BibitemOpen
  \bibfield  {author} {\bibinfo {author} {\bibfnamefont {A.-P.}\ \bibnamefont
  {Jauho}}, \bibinfo {author} {\bibfnamefont {N.~S.}\ \bibnamefont
  {Wingreen}},\ and\ \bibinfo {author} {\bibfnamefont {Y.}~\bibnamefont
  {Meir}},\ }\bibfield  {title} {\bibinfo {title} {Time-dependent transport in
  interacting and noninteracting resonant-tunneling systems},\ }\href@noop {}
  {\bibfield  {journal} {\bibinfo  {journal} {Physical Review B}\ }\textbf
  {\bibinfo {volume} {50}},\ \bibinfo {pages} {5528} (\bibinfo {year}
  {1994})}\BibitemShut {NoStop}%
\bibitem [{\citenamefont {Johnsen}\ and\ \citenamefont
  {Jauho}(1999)}]{keldysh3}%
  \BibitemOpen
  \bibfield  {author} {\bibinfo {author} {\bibfnamefont {K.}~\bibnamefont
  {Johnsen}}\ and\ \bibinfo {author} {\bibfnamefont {A.-P.}\ \bibnamefont
  {Jauho}},\ }\bibfield  {title} {\bibinfo {title} {Quasienergy spectroscopy of
  excitons},\ }\href@noop {} {\bibfield  {journal} {\bibinfo  {journal}
  {Physical review letters}\ }\textbf {\bibinfo {volume} {83}},\ \bibinfo
  {pages} {1207} (\bibinfo {year} {1999})}\BibitemShut {NoStop}%
\bibitem [{\citenamefont {Kamenev}(2023)}]{keldysh4}%
  \BibitemOpen
  \bibfield  {author} {\bibinfo {author} {\bibfnamefont {A.}~\bibnamefont
  {Kamenev}},\ }\href@noop {} {\emph {\bibinfo {title} {Field theory of
  non-equilibrium systems}}}\ (\bibinfo  {publisher} {Cambridge University
  Press},\ \bibinfo {year} {2023})\BibitemShut {NoStop}%
\bibitem [{\citenamefont {Shirley}(1965)}]{shirley_floquet}%
  \BibitemOpen
  \bibfield  {author} {\bibinfo {author} {\bibfnamefont {J.~H.}\ \bibnamefont
  {Shirley}},\ }\bibfield  {title} {\bibinfo {title} {Solution of the
  schr\"odinger equation with a hamiltonian periodic in time},\ }\href
  {https://doi.org/10.1103/PhysRev.138.B979} {\bibfield  {journal} {\bibinfo
  {journal} {Phys. Rev.}\ }\textbf {\bibinfo {volume} {138}},\ \bibinfo {pages}
  {B979} (\bibinfo {year} {1965})}\BibitemShut {NoStop}%
\bibitem [{Note2()}]{Note2}%
  \BibitemOpen
  \bibinfo {note} {In the situation where there is more than one
  Floquet-Lindblad steady state, $\protect \bm {G}$ becomes non-invertible and
  we will need to analyze its null space to obtain all the steady
  states.}\BibitemShut {Stop}%
\end{thebibliography}%

\end{document}